\documentclass[aip,reprint,author-year,nofootinbib]{revtex4-1}

\usepackage{graphicx}
\usepackage{amsmath,amssymb}             
\usepackage{stmaryrd}					
\usepackage{color}
\usepackage[dvipsnames]{xcolor}
\usepackage{color,hyperref}
\definecolor{darkblue}{rgb}{0.0,0.0,0.4}
\definecolor{red}{rgb}{0.7,0.0,0.0}
\definecolor{green}{rgb}{0.0,0.5,0.0}
\hypersetup{colorlinks,breaklinks,
            linkcolor=darkblue,urlcolor=darkblue,
            anchorcolor=darkblue,citecolor=RoyalBlue}

\usepackage{gensymb}
\usepackage[mathscr]{euscript}
\usepackage{upgreek}
\DeclareFontFamily{OT1}{pzc}{}
\DeclareFontShape{OT1}{pzc}{m}{it}{<-> s * [1.10] pzcmi7t}{}
\DeclareMathAlphabet{\mathpzc}{OT1}{pzc}{m}{it}
\usepackage{stmaryrd}
\usepackage{tikz}
\usetikzlibrary{shapes,arrows,shadows}
\renewcommand{\iint}{\int\!\!\!\!\!\int}
\usepackage{chngcntr}
\usepackage{ulem}

\def\apj{Astrophys. J.}

\def\apjs{Astrophys. J. Supp.}

\def\mnras{Mon. Not. R. Astron. Soc.}

\def\prd{Phys. Rev. D}

\def\nat{Nature}

\def\aap{Astron. \& Astrophys.}

\def\jcap{Journal of Cosmology and Astroparticle Physics}

\def\pasp{Publications of the Astronomical Society of the Pacific}

\newcommand{\comment}[1]{}

\newcommand{\bolfi}{\textsc{bolfi}}
\newcommand{\delfi}{\textsc{delfi}}
\newcommand{\p}{\mathpzc{P}}
\newcommand{\deltaM}{\delta\hspace{-0.1em}M}

\begin{document}


\title{Bayesian optimisation for likelihood-free cosmological inference}



\author{Florent Leclercq}
\email{florent.leclercq@polytechnique.org}
\homepage{http://www.florent-leclercq.eu/}
\affiliation{Imperial Centre for Inference and Cosmology (ICIC) \& Astrophysics Group, Imperial College London, Blackett Laboratory, Prince Consort Road, London SW7 2AZ, United Kingdom}


\date{\today}

\begin{abstract}
\noindent Many cosmological models have only a finite number of parameters of interest, but a very expensive data-generating process and an intractable likelihood function. We address the problem of performing likelihood-free Bayesian inference from such black-box simulation-based models, under the constraint of a very limited simulation budget (typically a few thousand). To do so, we adopt an approach based on the likelihood of an alternative parametric model. Conventional approaches to approximate Bayesian computation such as likelihood-free rejection sampling are impractical for the considered problem, due to the lack of knowledge about how the parameters affect the discrepancy between observed and simulated data. As a response, we make use of a strategy previously developed in the machine learning literature (Bayesian optimisation for likelihood-free inference, {\bolfi}), which combines Gaussian process regression of the discrepancy to build a surrogate surface with Bayesian optimisation to actively acquire training data. We extend the method by deriving an acquisition function tailored for the purpose of minimising the expected uncertainty in the approximate posterior density, in the parametric approach. The resulting algorithm is applied to the problems of summarising Gaussian signals and inferring cosmological parameters from the Joint Lightcurve Analysis supernovae data. We show that the number of required simulations is reduced by several orders of magnitude, and that the proposed acquisition function produces more accurate posterior approximations, as compared to common strategies.
\end{abstract}


\maketitle



\section{Introduction}
\label{sec:Introduction}

We consider the problem of Bayesian inference from cosmological data, in the common scenario where we can generate synthetic data through forward simulations, but where the exact likelihood function is intractable. The generative process can be extremely general: it may be a noisy non-linear dynamical system involving an unrestricted number of latent variables. Likelihood-free inference methods, also known as approximate Bayesian computation \citep[ABC, see][for reviews]{Marin2011,Lintusaari2017a} replace likelihood calculations with data model evaluations. In recent years, they have emerged as a viable alternative to likelihood-based techniques, when the simulator is sufficiently cheap. Applications in cosmology include measuring cosmological parameters from type Ia supernovae \citep{Weyant2013} and weak lensing peak counts \citep{Lin2015}, analysing the galaxy halo connection \citep{Hahn2017}, inferring the photometric and size evolution of galaxies \citep{Carassou2017}, measuring cosmological redshift distributions \citep{Kacprzak2018}, estimating the ionising background from the Lyman-$\alpha$ and Lyman-$\beta$ forests \citep{Davies2018}.

In its simplest form, ABC takes the form of likelihood-free rejection sampling and involves forward simulating data from parameters drawn from the prior, then accepting parameters when the discrepancy (by some measure) between simulated data and observed data is smaller than a user-specified threshold $\varepsilon$. Such an approach tends to be extremely expensive since many simulated data sets get rejected, due to the lack of knowledge about the relation between the model parameters and the corresponding discrepancy. Variants of likelihood-free rejection sampling such as Population (or Sequential) Monte Carlo ABC \citep[\textsc{pmc}-\textsc{abc} or \textsc{smc}-\textsc{abc}, see][for implementations aimed at astrophysical applications]{Akeret2015,Ishida2015,Jennings2017} improve upon this scheme by making the proposal adaptive; however, they do not use a probabilistic model for the relation between parameters and discrepancies (also known as a surrogate surface), so that their practical use usually necessitates $\mathcal{O}(10^4-10^6)$ evaluations of the simulator. 

In this paper, we address the challenging problem where the number of simulations is extremely limited, e.g. to a few thousand, rendering the use of sampling-based ABC methods impossible. To this end, we use Bayesian optimisation for likelihood-free inference \citep[{\bolfi},][]{GutmannCorander2016}, an algorithm which combines probabilistic modelling of the discrepancy with optimisation to facilitate likelihood-free inference. Since it was introduced, {\bolfi} has been applied to various statistical problems in science, including inference of the Ricker model \citep{GutmannCorander2016}, the Lotka-Volterra predator-prey model and population genetic models \citep{Jaervenpaeae2018}, pathogen spread models \citep{Lintusaari2017a}, atomistic structure models in materials \citep{Todorovic2017}, and cognitive models in human-computer interaction \citep{Kangasraeaesioe2017}. This work aims at introducing {\bolfi} in cosmological data analysis and at presenting its first cosmological application. We focus on computable parametric approximations to the true likelihood (also known as synthetic likelihoods), rendering the approach completely $\varepsilon$-free. Recently, \citet{Jaervenpaeae2017} introduced an acquisition function for Bayesian optimisation (the expected integrated variance), specifically tailored to perform efficient and accurate ABC. We extend their work by deriving the expression of the expected integrated variance in the parametric approach. This acquisition function measures the expected uncertainty in the estimate of the {\bolfi} posterior density, which is due to the limited number of simulations, over the future evaluation of the simulation model. The next simulation location is proposed so that this expected uncertainty is minimised. As a result, high-fidelity posterior inferences can be obtained with orders of magnitude fewer simulations than with likelihood-free rejection sampling. As examples, we demonstrate the use of {\bolfi} on the problems of summarising Gaussian signals and inferring cosmological parameters from the Joint Lightcurve Analysis (JLA) supernovae data set \citep{Betoule2014}.

The structure of this paper is as follows. In section \ref{sec:Inference of simulator-based statistical models}, we provide a review of the formalism for the inference of simulator-based statistical models. In section \ref{sec:Regression and Optimisation for likelihood-free inference}, we describe {\bolfi} and discuss the regression and optimisation strategies. In particular, we provide the optimal acquisition rule for ABC in the parametric approach to likelihood approximation. Applications are given in section \ref{sec:Applications}. The developed method is discussed in section \ref{sec:Discussion} in the context of cosmological data analysis. Section \ref{sec:Conclusion} concludes the paper. Mathematical details and descriptions of the case studies are presented in the appendices. 

\section{Inference of simulator-based statistical models}
\label{sec:Inference of simulator-based statistical models}

\subsection{Simulator-based statistical models}
\label{ssec:Simulator-based statistical models}

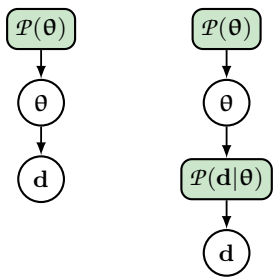
\begin{figure}[h]
\begin{center}
\begin{tikzpicture}
	\pgfdeclarelayer{background}
	\pgfdeclarelayer{foreground}
	\pgfsetlayers{background,main,foreground}

	\tikzstyle{probability}=[draw, thick, text centered, rounded corners, minimum height=1em, minimum width=1em, fill=green!20]
	\tikzstyle{variabl}=[draw, thick, text centered, circle, minimum height=1em, minimum width=1em]

	\def\blockdist{0.7}
	\def\modeldist{2.0}

    \node (thetaprobaii) [probability]
    {$\p(\boldsymbol{\uptheta})$};
    \path (thetaprobaii.south)+(0,-\blockdist) node (thetaii) [variabl]
    {$\boldsymbol{\uptheta}$};
    \path (thetaii.south)+(0,-\blockdist) node (dprobaii) [probability]
    {$\p(\textbf{d}|\boldsymbol{\uptheta})$};
    \path (dprobaii.south)+(0,-\blockdist) node (dii) [variabl]
    {$\textbf{d}$};
    
    \path (thetaprobaii.west)+(-\modeldist,0) node (thetaprobai) [probability]
    {$\p(\boldsymbol{\uptheta})$};
    \path (thetaprobai.south)+(0,-\blockdist) node (thetai) [variabl]
    {$\boldsymbol{\uptheta}$};
    \path (thetai.south)+(0,-\blockdist) node (di) [variabl]
    {$\textbf{d}$};

	\path [draw, line width=0.7pt, arrows={-latex}] (thetaprobaii) -- (thetaii);
	\path [draw, line width=0.7pt, arrows={-latex}] (thetaii) -- (dprobaii);
	\path [draw, line width=0.7pt, arrows={-latex}] (dprobaii) -- (dii);

	\path [draw, line width=0.7pt, arrows={-latex}] (thetaprobai) -- (thetai);
	\path [draw, line width=0.7pt, arrows={-latex}] (thetai) -- (di);

\end{tikzpicture}
\end{center}
\caption{Hierarchical representation of the exact Bayesian problem for simulator-based statistical models of different complexities: a deterministic simulator (left), and a stochastic simulator (right).\label{fig:BHM_exact}}
\end{figure}

Simulator-based statistical models (also known as generative models) can be written in a hierarchical form (figure \ref{fig:BHM_exact}), where $\boldsymbol{\uptheta}$ are the parameters of interest, and $\textbf{d}$ the simulated data. $\p(\boldsymbol{\uptheta})$ is the prior probability distribution of $\boldsymbol{\uptheta}$ and $\p(\textbf{d}|\boldsymbol{\uptheta})$ is the sampling distribution of $\textbf{d}$ given $\boldsymbol{\uptheta}$.

The simplest case (figure \ref{fig:BHM_exact}, left) is when the simulator is a deterministic function of its input and does not use any random variable, i.e.
\begin{equation}
\p(\textbf{d}|\boldsymbol{\uptheta}) = \updelta_\mathrm{D}(\textbf{d} - \boldsymbol{\hat{\mathrm{d}}}(\boldsymbol{\uptheta})) ,\label{eq:Dirac_deterministic_simulator}
\end{equation}
where $\updelta_\mathrm{D}$ is a Dirac delta distribution and $\boldsymbol{\hat{\mathrm{d}}}$ a deterministic function of $\boldsymbol{\uptheta}$.

In a more generic scenario (figure \ref{fig:BHM_exact}, right), the simulator is stochastic, in the sense that the data are drawn from an overall (but often unknown analytically) probability distribution function (pdf) $\p(\textbf{d}|\boldsymbol{\uptheta})$. Equation \eqref{eq:Dirac_deterministic_simulator} does not hold in this case. The scatter between different realisations of $\textbf{d}$ given the same $\boldsymbol{\uptheta}$ can have various origins. In the simplest case, it only reflects the intrinsic uncertainty, which is of interest. More generically, additional nuisance parameters can be at play to produce the data $\textbf{d}$ and will contribute to the uncertainty. This ``latent space'' can often be hundred-to-multi-million dimensional. Simulator-based cosmological models are typically of this kind: although the physical and observational processes simulated are repeatable features about which inferences can be made, the particular realisation of Fourier phases of the data is entirely noise-driven. Ideally, phase-dependent quantities should not contribute to any measure of match or mismatch between model and data.

\subsection{The exact Bayesian problem}
\label{ssec:The exact Bayesian problem}

The inference problem is to evaluate the probability of $\boldsymbol{\uptheta}$ given $\textbf{d}$,
\begin{equation}
\p(\boldsymbol{\uptheta}|\textbf{d}) = \p(\textbf{d}|\boldsymbol{\uptheta}) \, \frac{\p(\boldsymbol{\uptheta})}{\p(\textbf{d})},
\label{eq:exact_problem_Bayes}
\end{equation}
for the observed data $\textbf{d}_\mathrm{O}$, i.e.
\begin{equation}
\p(\boldsymbol{\uptheta}|\textbf{d})_\mathrm{|\textbf{d}=\textbf{d}_O} = \mathcal{L}(\boldsymbol\uptheta) \, \frac{\p(\boldsymbol{\uptheta})}{Z_\textbf{d}} ,
\end{equation}
where the exact likelihood for the problem is defined as
\begin{equation}
\mathcal{L}(\boldsymbol\uptheta) \equiv \p(\textbf{d}|\boldsymbol\uptheta)_\mathrm{|\textbf{d}=\textbf{d}_O} .
\end{equation}
It is generally of unknown analytical form. The normalisation constant is $Z_\textbf{d} \equiv \p(\textbf{d})_\mathrm{|\textbf{d}=\textbf{d}_O}$, where $\p(\textbf{d})$ is the marginal distribution of $\textbf{d}$.

\subsection{Approximate Bayesian computation}
\label{ssec:Approximate Bayesian computation}

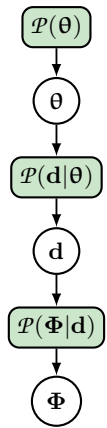
\begin{figure}[h]
\begin{center}
\begin{tikzpicture}
	\pgfdeclarelayer{background}
	\pgfdeclarelayer{foreground}
	\pgfsetlayers{background,main,foreground}

	\tikzstyle{probability}=[draw, thick, text centered, rounded corners, minimum height=1em, minimum width=1em, fill=green!20]
	\tikzstyle{variabl}=[draw, thick, text centered, circle, minimum height=1em, minimum width=1em]

	\def\blockdist{0.7}

    \node (thetaproba) [probability]
    {$\p(\boldsymbol{\uptheta})$};
    \path (thetaproba.south)+(0,-\blockdist) node (theta) [variabl]
    {$\boldsymbol{\uptheta}$};
    \path (theta.south)+(0,-\blockdist) node (dproba) [probability]
    {$\p(\textbf{d}|\boldsymbol{\uptheta})$};
    \path (dproba.south)+(0,-\blockdist) node (d) [variabl]
    {$\textbf{d}$};
    \path (d.south)+(0,-\blockdist) node (phiproba) [probability]
    {$\p(\boldsymbol{\Phi}|\textbf{d})$};
    \path (phiproba.south)+(0,-\blockdist) node (phi) [variabl]
    {$\boldsymbol{\Phi}$};

	\path [draw, line width=0.7pt, arrows={-latex}] (thetaproba) -- (theta);
	\path [draw, line width=0.7pt, arrows={-latex}] (theta) -- (dproba);
	\path [draw, line width=0.7pt, arrows={-latex}] (dproba) -- (d);
	\path [draw, line width=0.7pt, arrows={-latex}] (d) -- (phiproba);
	\path [draw, line width=0.7pt, arrows={-latex}] (phiproba) -- (phi);

\end{tikzpicture}
\end{center}
\caption{Hierarchical representation of the approximate Bayesian inference problem for simulator-based statistical models, with a compression of the raw data to a set of summary statistics.\label{fig:BHM_approx}}
\end{figure}

Inference of simulator-based statistical models is usually based on a finite set of simulated data $\textbf{d}_{\boldsymbol{\uptheta}}$, generated with parameter value $\boldsymbol{\uptheta}$, and on a measurement of the discrepancy between simulated data and observed data $\textbf{d}_\mathrm{O}$. This discrepancy is used to define an approximation to the exact likelihood $\mathcal{L}(\boldsymbol{\uptheta})$. The approximation happens on multiple levels.

On a physical and statistical level, the approximation consists of compressing the full data $\textbf{d}_\mathrm{O}$ to a set of summary statistics $\boldsymbol{\Phi}_\mathrm{O}$ before performing inference. Similarly, simulated data $\textbf{d}_{\boldsymbol{\uptheta}}$ are compressed to simulated summary statistics $\boldsymbol{\Phi}_{\boldsymbol{\uptheta}}$. This can be seen as adding a layer to the Bayesian hierarchical model (figure \ref{fig:BHM_approx}). The purpose of this operation is to filter out the information in $\textbf{d}$ that is not deemed relevant to the inference of $\boldsymbol{\uptheta}$, so as to reduce the dimensionality of the problem. Ideally, $\boldsymbol{\Phi}$ should be \textit{sufficient} for parameters $\boldsymbol{\uptheta}$, i.e. formally $\p(\boldsymbol{\uptheta}|\boldsymbol{\Phi}) = \p(\boldsymbol{\uptheta}|\boldsymbol{\Phi},\textbf{d})$ or equivalently $\p(\textbf{d}|\boldsymbol{\Phi},\boldsymbol{\uptheta}) = \p(\textbf{d}|\boldsymbol{\Phi})$, which happens when the compression is lossless. However, sufficient summary statistics are generally unknown or even impossible to design; therefore the compression from $\textbf{d}$ to $\boldsymbol{\Phi}$ will usually be lossy. The approximate inference problem to be solved is now $\p(\boldsymbol{\uptheta}|\boldsymbol{\Phi}) = \p(\boldsymbol{\Phi}|\boldsymbol{\uptheta}) \, \dfrac{\p(\boldsymbol{\uptheta})}{\p(\boldsymbol{\Phi})}$ for the observed summary statistics $\boldsymbol{\Phi}_\mathrm{O}$, i.e.
\begin{equation}
\p(\boldsymbol{\uptheta}|\boldsymbol{\Phi})_\mathrm{|\boldsymbol{\Phi}=\boldsymbol{\Phi}_O} = L(\boldsymbol\uptheta) \, \frac{\p(\boldsymbol{\uptheta})}{Z_{\boldsymbol{\Phi}}} .
\label{eq:approx_problem_Bayes}
\end{equation}
In other words, $\mathcal{L}(\boldsymbol{\uptheta})$ is replaced by
\begin{equation}
L(\boldsymbol{\uptheta}) \equiv \p(\boldsymbol\Phi|\boldsymbol\uptheta)_\mathrm{|\boldsymbol{\Phi}=\boldsymbol{\Phi}_O} ,
\label{eq:L_theta}
\end{equation}
and $Z_\textbf{d}$ by $Z_{\boldsymbol{\Phi}} \equiv \p(\boldsymbol{\Phi})_{|\boldsymbol{\Phi}=\boldsymbol{\Phi}_\mathrm{O}}$. Inference of model \ref{fig:BHM_approx} gives
\begin{equation}
\p(\boldsymbol\uptheta, \textbf{d} | \boldsymbol\Phi) \propto \p(\boldsymbol\Phi|\textbf{d}) \, \p(\textbf{d}|\boldsymbol\uptheta) \, \p(\boldsymbol\uptheta),
\label{eq:BHM_approx_expansion}
\end{equation}
with, after marginalisation over $\textbf{d}$,
\begin{equation}
\p(\boldsymbol\uptheta | \boldsymbol\Phi) = \int \p(\boldsymbol\uptheta, \textbf{d} | \boldsymbol\Phi) \, \mathrm{d}\textbf{d} .
\label{eq:BHM_approx_marginalisation}
\end{equation}
Therefore, the approximate likelihood $L(\boldsymbol{\uptheta})$ must satisfy
\begin{equation}
L(\boldsymbol{\uptheta}) \propto \int \p(\boldsymbol\Phi|\textbf{d})_\mathrm{|\boldsymbol{\Phi}=\boldsymbol{\Phi}_O} \, \p(\textbf{d}|\boldsymbol\uptheta) \, \mathrm{d}\textbf{d} .
\label{eq:BHM_approx_likelihood}
\end{equation}
In many cases, the compression from $\textbf{d}$ to $\boldsymbol{\Phi}$ is deterministic, i.e.
\begin{equation}
\p(\boldsymbol{\Phi}|\textbf{d}) = \updelta_\mathrm{D}(\boldsymbol{\Phi} - \boldsymbol{\hat{\Phi}}(\textbf{d})) ,
\label{eq:Dirac_compression}
\end{equation}
which simplifies the integral over $\textbf{d}$ in equations \eqref{eq:BHM_approx_marginalisation} and \eqref{eq:BHM_approx_likelihood}.

On a practical level, $L(\boldsymbol{\uptheta})$ is still of unknown analytical form (which is a property of $\p(\boldsymbol\Phi|\boldsymbol\uptheta)$ inherited from $\p(\textbf{d}|\boldsymbol{\uptheta})$ in model \ref{fig:BHM_approx}). Therefore, it has to be approximated using the simulator. We denote by $\widehat{L}^N(\boldsymbol{\uptheta})$ an estimate of $L(\boldsymbol{\uptheta})$ computed using $N$ realisations of the simulator. The limiting approximation, in the case where infinite computer resources were available, is denoted by $\widetilde{L}(\boldsymbol{\uptheta})$, such that
\begin{equation}
\widehat{L}^N(\boldsymbol{\uptheta}) \xrightarrow[N \rightarrow \infty]{} \widetilde{L}(\boldsymbol{\uptheta}) .
\end{equation}
Note that $\widetilde{L}(\boldsymbol{\uptheta})$ can be different from $L(\boldsymbol{\uptheta})$, depending on the assumptions made to construct $\widehat{L}^N(\boldsymbol{\uptheta})$. These are discussed in section \ref{ssec:Computable approximations of the likelihood}.

\subsection{Computable approximations of the likelihood}
\label{ssec:Computable approximations of the likelihood}

\subsubsection{Deterministic simulators}
\label{sssec:Deterministic simulators}

The simplest possible case is when the simulator does not use any random variable, i.e. $\boldsymbol{\Phi}_{\boldsymbol{\uptheta}}$ is an entirely deterministic function of $\boldsymbol{\uptheta}$ (see figure \ref{fig:BHM_exact}, left). Equivalently, all the conditional probabilities appearing in equation \eqref{eq:BHM_approx_expansion} reduce to Dirac delta distributions given by equations \eqref{eq:Dirac_deterministic_simulator} and \eqref{eq:Dirac_compression}. In this case, one can directly use the approximate likelihood given by equation \eqref{eq:L_theta}, complemented by an assumption on the functional shape of $\p(\boldsymbol{\Phi}|\boldsymbol{\uptheta})$.

\subsubsection{Parametric approximations and the synthetic likelihood}
\label{sssec:Parametric approximations and the synthetic likelihood}

When the simulator is not deterministic, the pdf $\p(\boldsymbol{\Phi}|\boldsymbol{\uptheta})$ is unknown analytically. Nonetheless, in some situations, it may be reasonably assumed to follow specific parametric forms.

For example, if $\boldsymbol{\Phi}_{\boldsymbol{\uptheta}}$ is obtained through averaging a sufficient number of independent and identically distributed variables contained in $\textbf{d}$, the central limit theorem suggests that a Gaussian distribution is appropriate, i.e. $\widetilde{L}(\boldsymbol{\uptheta}) = \exp\left[\tilde{\ell}(\boldsymbol{\uptheta})\right]$ with 
\begin{equation}
-2 \tilde{\ell}(\boldsymbol{\uptheta}) \equiv \log \left| 2\pi \boldsymbol{\Sigma}_{\boldsymbol{\uptheta}} \right| + (\boldsymbol{\Phi}_\mathrm{O} - \boldsymbol{\upmu}_{\boldsymbol{\uptheta}})^\intercal \boldsymbol{\Sigma}_{\boldsymbol{\uptheta}}^{-1} (\boldsymbol{\Phi}_\mathrm{O} - \boldsymbol{\upmu}_{\boldsymbol{\uptheta}}),
\end{equation}
where the mean and covariance matrix,
\begin{equation}
\boldsymbol{\upmu}_{\boldsymbol{\uptheta}} \equiv \mathrm{E}\left[ \boldsymbol{\Phi}_{\boldsymbol{\uptheta}} \right] \enskip \mathrm{and} \enskip \boldsymbol{\Sigma}_{\boldsymbol{\uptheta}} \equiv \mathrm{E}\left[ (\boldsymbol{\Phi}_{\boldsymbol{\uptheta}}-\boldsymbol{\upmu}_{\boldsymbol{\uptheta}}) (\boldsymbol{\Phi}_{\boldsymbol{\uptheta}}-\boldsymbol{\upmu}_{\boldsymbol{\uptheta}})^\intercal \right],
\end{equation}
can depend on $\boldsymbol{\uptheta}$. This is an approximation of $L(\boldsymbol{\uptheta})$, unless the summary statistics $\boldsymbol{\Phi}_{\boldsymbol{\uptheta}}$ are indeed Gaussian-distributed. $\boldsymbol{\upmu}_{\boldsymbol{\uptheta}}$ and $\boldsymbol{\Sigma}_{\boldsymbol{\uptheta}}$ are generally unknown, but can be estimated using the simulator: given a set of $N$ simulations $\lbrace \boldsymbol{\Phi}_{\boldsymbol{\uptheta}}^{(i)} \rbrace$, drawn independently from $\p(\boldsymbol{\Phi}|\boldsymbol{\uptheta})$, one can define
\begin{equation}
\boldsymbol{\hat{\upmu}}_{\boldsymbol{\uptheta}} \equiv \mathrm{E}^N\left[ \boldsymbol{\Phi}_{\boldsymbol{\uptheta}} \right] \enskip \mathrm{and} \enskip \boldsymbol{\hat{\Sigma}}_{\boldsymbol{\uptheta}} \equiv \mathrm{E}^N\left[ (\boldsymbol{\Phi}_{\boldsymbol{\uptheta}}-\boldsymbol{\hat{\upmu}}_{\boldsymbol{\uptheta}}) (\boldsymbol{\Phi}_{\boldsymbol{\uptheta}}-\boldsymbol{\hat{\upmu}}_{\boldsymbol{\uptheta}})^\intercal \right],
\label{eq:mean_covariance_empirical}
\end{equation}
where $\mathrm{E}^N$ stands for the empirical average over the set of simulations. A computable approximation of the likelihood is therefore $\widehat{L}^N(\boldsymbol{\uptheta}) = \exp\left[ \hat{\ell}^N(\boldsymbol{\uptheta}) \right]$, where
\begin{equation}
-2 \hat{\ell}^N(\boldsymbol{\uptheta}) \equiv \log \left| 2\pi \boldsymbol{\hat{\Sigma}}_{\boldsymbol{\uptheta}} \right| + (\boldsymbol{\Phi}_\mathrm{O} - \boldsymbol{\hat{\upmu}}_{\boldsymbol{\uptheta}})^\intercal \boldsymbol{\hat{\Sigma}}_{\boldsymbol{\uptheta}}^{-1} (\boldsymbol{\Phi}_\mathrm{O} - \boldsymbol{\hat{\upmu}}_{\boldsymbol{\uptheta}}).
\label{eq:synthetic_likelihood}
\end{equation}
Due to the approximation of the expectation $\mathrm{E}$ with an empirical average $\mathrm{E}^N$, both $\boldsymbol{\hat{\upmu}}_{\boldsymbol{\uptheta}}$ and $\boldsymbol{\hat{\Sigma}}_{\boldsymbol{\uptheta}}$ become random objects. The approximation of the likelihood $\widehat{L}^N(\boldsymbol{\uptheta})$ is therefore a random function with some intrinsic uncertainty itself, and its computation is a stochastic process. This is further discussed using a simple example in section \ref{ssec:Summarising Gaussian signals}.

The approximation given in equation \eqref{eq:synthetic_likelihood}, known as the synthetic likelihood \citep{Wood2010,Price2017}, has already been applied successfully to perform approximate inference in several scientific fields. However, as pointed out by \citet{SellentinHeavens2016}, for inference from Gaussian-distributed summaries $\boldsymbol{\Phi}_{\boldsymbol{\uptheta}}$ with an estimated covariance matrix $\boldsymbol{\hat{\Sigma}}_{\boldsymbol{\uptheta}}$, a different parametric form, namely a multivariate $t$-distribution, should rather be used. The investigation of a synthetic $t$-likelihood is left to future investigations.

In section \ref{ssec:Summarising Gaussian signals} and appendix \ref{apx:Summarising Gaussian signals}, we extend previous work on the Gaussian synthetic likelihood and introduce a Gamma synthetic likelihood for case where the $\boldsymbol{\Phi}_{\boldsymbol{\uptheta}}$ are (or can be assumed to be) Gamma-distributed. 

\subsubsection{Non-parametric approximations and likelihood-free rejection sampling}
\label{sssec:Non-parametric approximations and likelihood-free rejection sampling}

An alternative to assuming a parametric form for $L(\boldsymbol{\uptheta})$ is to replace it by a kernel density estimate of the distribution of a discrepancy between simulated and observed summary statistics, i.e.
\begin{equation}
\widetilde{L}(\boldsymbol{\uptheta}) \equiv \mathrm{E}\left[ \kappa(\Delta_{\boldsymbol{\uptheta}}) \right],
\end{equation}
where $\Delta_{\boldsymbol{\uptheta}}$ is a non-negative function of $\boldsymbol{\Phi}_\mathrm{O}$ and $\boldsymbol{\Phi}_{\boldsymbol{\uptheta}}$ (usually of $\boldsymbol{\Phi}_\mathrm{O} - \boldsymbol{\Phi}_{\boldsymbol{\uptheta}}$) which can also possibly depend on $\boldsymbol{\uptheta}$ and any variable used internally by the simulator, and the kernel $\kappa$ is a non-negative, univariate function independent of $\boldsymbol{\uptheta}$ (usually with a maximum at zero). A computable approximation of the likelihood is then given by
\begin{equation}
\widehat{L}^N(\boldsymbol{\uptheta}) \equiv \mathrm{E}^N\left[ \kappa(\Delta_{\boldsymbol{\uptheta}}) \right] .
\end{equation}

For likelihood-free inference, $\kappa$ is often chosen as the uniform kernel on the interval $\left[ 0, \varepsilon \right)$, i.e. $\kappa(u) \propto \chi_{\left[ 0, \varepsilon \right)}(u)$, where $\varepsilon$ is called the threshold and the indicator function $\chi_{\left[ 0, \varepsilon \right)}$ equals one if $u \in \left[ 0, \varepsilon \right)$ and zero otherwise. This yields
\begin{equation}
\widetilde{L}(\boldsymbol{\uptheta}) \propto \p(\Delta_{\boldsymbol{\uptheta}} \leq \varepsilon) \quad \mathrm{and} \quad \widehat{L}^N(\boldsymbol{\uptheta}) \propto \p^N(\Delta_{\boldsymbol{\uptheta}} \leq \varepsilon),
\label{eq:approximate_likelihood_acceptance}
\end{equation}
where $\p^N(\Delta_{\boldsymbol{\uptheta}} \leq \varepsilon)$ is the empirical probability that the discrepancy is below the threshold. $\widehat{L}^N(\boldsymbol{\uptheta})$ can be straightforwardly evaluated by running simulations, computing $\Delta_{\boldsymbol{\uptheta}}$ and using $\Delta_{\boldsymbol{\uptheta}} \leq \varepsilon$ as a criterion for acceptance or rejection of proposed samples. Such an approach is often simply (or mistakenly) referred to as approximate Bayesian computation (ABC) in the astrophysics literature, although the more appropriate and explicit denomination is likelihood-free rejection sampling \citep[see e.g.][]{Marin2011}.

It is interesting to note that the parametric approximate likelihood approach of section \ref{sssec:Parametric approximations and the synthetic likelihood} can be embedded into the non-parametric approach. Indeed, $\Delta_{\boldsymbol{\uptheta}}$ can be defined as
\begin{equation}
\Delta^{\textbf{C}_{\boldsymbol{\uptheta}}}_{\boldsymbol{\uptheta}} \equiv \log|2\pi \textbf{C}_{\boldsymbol{\uptheta}}| + (\boldsymbol{\Phi}_\mathrm{O} - \boldsymbol{\Phi}_{\boldsymbol{\uptheta}})^\intercal \textbf{C}_{\boldsymbol{\uptheta}}^{-1} (\boldsymbol{\Phi}_\mathrm{O} - \boldsymbol{\Phi}_{\boldsymbol{\uptheta}})
\end{equation}
for some positive semidefinite matrix $\textbf{C}_{\boldsymbol{\uptheta}}$. The second term is the square of the Mahalanobis distance, which includes the Euclidean distance as a special case, when $\textbf{C}_{\boldsymbol{\uptheta}}$ is the identity matrix. Using an exponential kernel $\kappa(u) = \exp(-u/2)$ and $\textbf{C}_{\boldsymbol{\uptheta}} = \boldsymbol{\hat{\Sigma}}_{\boldsymbol{\uptheta}}$ gives $\widetilde{L}(\boldsymbol{\uptheta}) = \mathrm{E}\left[ \kappa(\Delta_{\boldsymbol{\uptheta}}^{\boldsymbol{\hat{\Sigma}}_{\boldsymbol{\uptheta}}}) \right]$ and $\widehat{L}^N(\boldsymbol{\uptheta}) = \mathrm{E}^N\left[ \kappa(\Delta_{\boldsymbol{\uptheta}}^{\boldsymbol{\hat{\Sigma}}_{\boldsymbol{\uptheta}}}) \right]$ with
\begin{eqnarray}
-2 \log\left[\kappa(\Delta_{\boldsymbol{\uptheta}}^{\boldsymbol{\hat{\Sigma}}_{\boldsymbol{\uptheta}}}) \right] & = & \log \left| 2\pi \boldsymbol{\hat{\Sigma}}_{\boldsymbol{\uptheta}} \right| \\
& & + (\boldsymbol{\Phi}_\mathrm{O} - \boldsymbol{\Phi}_{\boldsymbol{\uptheta}})^\intercal \boldsymbol{\hat{\Sigma}}_{\boldsymbol{\uptheta}}^{-1} (\boldsymbol{\Phi}_\mathrm{O} - \boldsymbol{\Phi}_{\boldsymbol{\uptheta}}), \nonumber
\end{eqnarray}
the form of which is similar to equation \eqref{eq:synthetic_likelihood}. In fact, \citet[][proposition 1]{GutmannCorander2016} show that the synthetic likelihood satisfies
\begin{eqnarray}
-2\tilde{\ell}(\boldsymbol{\uptheta}) & = & J(\boldsymbol{\uptheta}) + \mathrm{constant}, \quad \mathrm{and}\label{eq:l_J_proposition_1}\\
-2\hat{\ell}^N(\boldsymbol{\uptheta}) & = & \widehat{J}^N(\boldsymbol{\uptheta}) + \mathrm{constant},\label{eq:l_J_proposition_2}
\end{eqnarray}
where 
\begin{equation}
J(\boldsymbol{\uptheta}) \equiv \mathrm{E}\left[ \Delta_{\boldsymbol{\uptheta}}^{\textbf{C}_{\boldsymbol{\uptheta}}} \right]
\label{eq:def_J}
\end{equation}
and
\begin{equation}
\widehat{J}^N(\boldsymbol{\uptheta}) \equiv \mathrm{E}^N\left[ \Delta_{\boldsymbol{\uptheta}}^{\textbf{C}_{\boldsymbol{\uptheta}}} \right]
\label{eq:def_J_N}
\end{equation}
are respectively the expectation and the empirical average of the discrepancy $\Delta_{\boldsymbol{\uptheta}}^{\textbf{C}_{\boldsymbol{\uptheta}}}$, for $\textbf{C}_{\boldsymbol{\uptheta}}= \boldsymbol{\hat{\Sigma}}_{\boldsymbol{\uptheta}}$.

\section{Regression and Optimisation for likelihood-free inference}
\label{sec:Regression and Optimisation for likelihood-free inference}

\subsection{Computational difficulties with likelihood-free rejection sampling}
\label{ssec:Computational difficulties with likelihood-free rejection sampling}

We have seen in section \ref{ssec:Computable approximations of the likelihood} that computable approximations $\widehat{L}^N(\boldsymbol{\uptheta})$ of the likelihood $L(\boldsymbol{\uptheta})$ are stochastic processes, due to the use of simulations to approximate intractable expectations. In the most popular ABC approach, i.e. likelihood-free rejection sampling (see section \ref{sssec:Non-parametric approximations and likelihood-free rejection sampling}), the expectations are approximated by empirical probabilities that the discrepancy is below the threshold $\varepsilon$. While this approach allows inference of simulator-based statistical models with minimal assumptions, it suffers from several limitations that can make its use impossible in practice.
\begin{enumerate}
\item It rejects most of the proposed samples when $\varepsilon$ is small, leading to a computationally inefficient algorithm.
\item It does not make assumptions about the shape or smoothness of the target function $L(\boldsymbol{\uptheta})$, hence accepted samples cannot ``share'' information in parameter space.
\item It uses a fixed proposal distribution (typically the prior $\p(\boldsymbol{\uptheta})$) and does not make use of already accepted samples to update the proposal of new points.
\item It aims at equal accuracy for all regions in parameter space, regardless of the values of the likelihood.
\end{enumerate}

To overcome these issues, the proposed approach follows closely \citet{GutmannCorander2016}, who combine regression of the discrepancy (addressing issues 1 and 2) with Bayesian optimisation (addressing issues 3 and 4) in order to improve the computational efficiency of inference of simulator-based models. In this work, we focus on parametric approximations of the likelihood; we refer to \citet{GutmannCorander2016} for a treatment of the non-parametric approach.

\subsection{Regression of the discrepancy}
\label{ssec:Regression of the discrepancy}

\begin{figure}
\begin{center}
\includegraphics[width=\columnwidth]{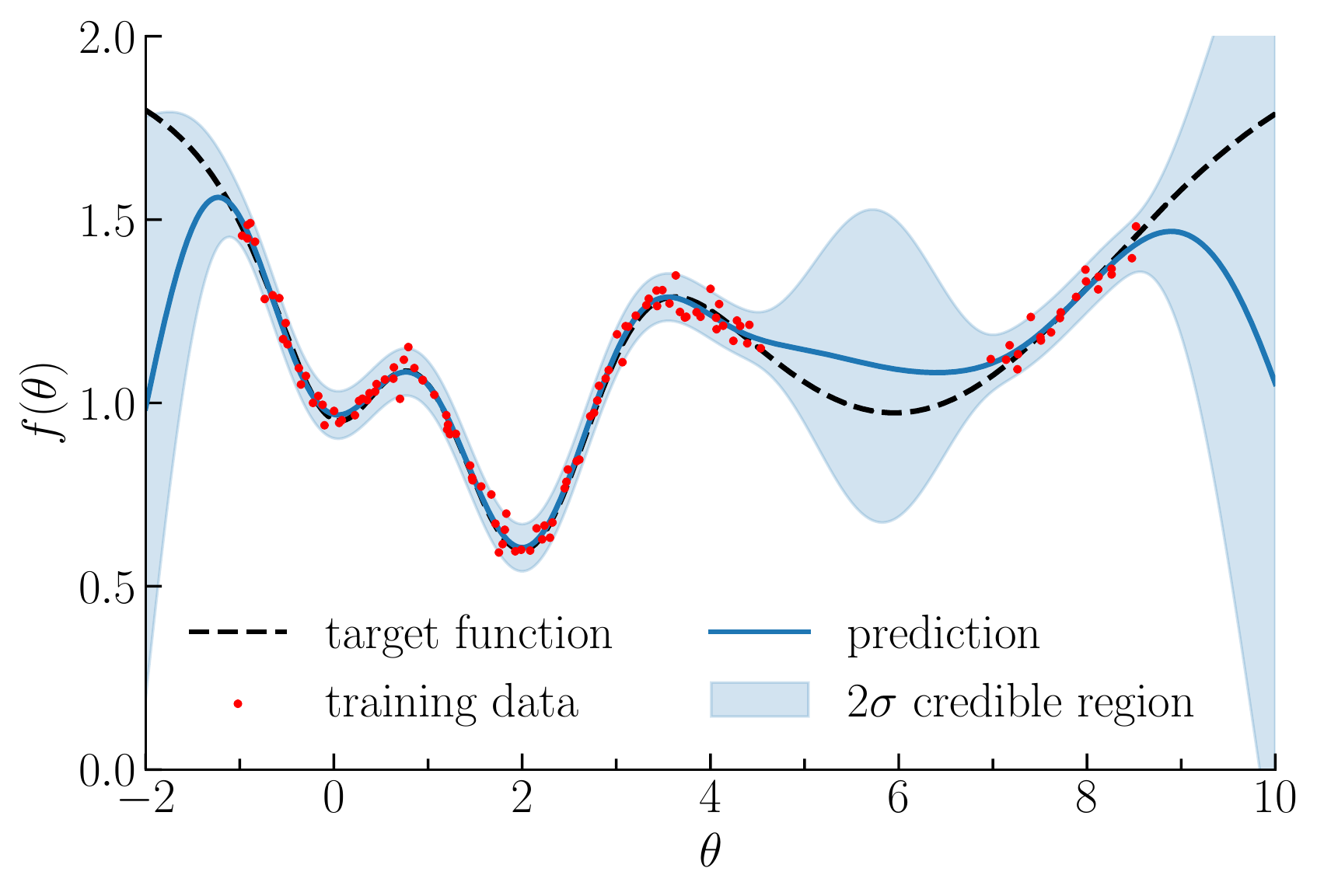} 
\caption{Illustration of Gaussian process regression in one dimension, for the target test function $f: \theta \mapsto 2 - \exp\left[-(\theta - 2)^2\right] - \exp\left[-(\theta - 6)^2/10\right] - 1/ (\theta^2 + 1)$ (dashed line). Training data are acquired (red dots); they are subject to a Gaussian observation noise with standard deviation $\sigma_\mathrm{n} = 0.03$. The blue line shows the mean prediction $\mu(\theta)$ of the Gaussian process regression, and the shaded region the corresponding $2\sigma(\theta)$ uncertainty. Gaussian processes allow interpolating and extrapolating predictions in regions of parameter space where training data are absent.\label{fig:GP_illustration}}
\end{center}
\end{figure}

The standard approach to obtain a computable approximate likelihood relies on empirical averages (equations \eqref{eq:mean_covariance_empirical} and \eqref{eq:def_J_N}). However, such sample averages are not the only way to approximate intractable expectations. Equations \eqref{eq:l_J_proposition_1} and \eqref{eq:def_J} show that, up to constants and the sign, $\tilde{\ell}(\boldsymbol{\uptheta})$ can be interpreted as a regression function with the model parameters $\boldsymbol{\uptheta}$ (the ``predictors'') as the independent input variables and the discrepancy $\Delta_{\boldsymbol{\uptheta}}$ as the response variable. Therefore, in the present approach, we consider an approximation of the intractable expectation defining $J(\boldsymbol{\uptheta})$ in equation \eqref{eq:def_J} based on a regression analysis of $\Delta_{\boldsymbol{\uptheta}}$, instead of sample averages. Explicitly, we consider
\begin{equation}
\widehat{J}^{(\mathrm{t})}(\boldsymbol{\uptheta}) \equiv \mathrm{E}^{(\mathrm{t})}\left[ \Delta_{\boldsymbol{\uptheta}}^{\textbf{C}_{\boldsymbol{\uptheta}}} \right],
\label{eq:def_J_t}
\end{equation}
where the superscript $(\mathrm{t})$ stands for ``training'' and the expectation $\mathrm{E}^{(\mathrm{t})}$ is taken under the probabilistic model defined in the following.

Inferring $J(\boldsymbol{\uptheta})$ via regression requires a training data set $\lbrace (\boldsymbol{\uptheta}^{(i)} , \Delta_{\boldsymbol{\uptheta}}^{(i)}) \rbrace\vspace*{-2pt}$ where the discrepancies are computed from the simulated summary statistics $\boldsymbol{\Phi}_{\boldsymbol{\uptheta}}^{(i)}$. Building this training set requires to run simulations, but does not involve an accept/reject criterion as does likelihood-free rejection sampling (thus addressing issue 1, see section \ref{ssec:Computational difficulties with likelihood-free rejection sampling}). A regression-based approach also allows incorporating a smoothness assumption about $J(\boldsymbol{\uptheta})$. In this way, samples of the training set can ``share'' the information of the computed $\Delta_{\boldsymbol{\uptheta}}$ in the neighbourhood of $\boldsymbol{\uptheta}$ (thus addressing issue 2). This suggests that fewer simulated data are needed to reach a certain level of accuracy when learning the target function $J(\boldsymbol{\uptheta})$.

In this work, we rely on Gaussian process (GP) regression in order to construct a prediction for $J(\boldsymbol{\uptheta})$. There are several reasons why this choice is advantageous for likelihood-free inference. First, GPs are a general-purpose regressor, able to deal with a large variety of functional shapes for $J(\boldsymbol{\uptheta})$, including potentially complex non-linear, or multi-modal features. Second, GPs provide not only a prediction (the mean of the regressed function), but also the uncertainty of the regression. This is useful for actively constructing the training data via Bayesian optimisation, as we show in section \ref{ssec:Acquisition rules}. Finally, GPs allow extrapolating the prediction into regions of the parameter space where no training points are available. These three properties are shown in figure \ref{fig:GP_illustration} for a multi-modal test function subject to observation noise.

We now briefly review Gaussian process regression. Suppose that we have a set of $t$ training points, $(\boldsymbol{\Theta}, \textbf{f}) \equiv \lbrace (\boldsymbol{\uptheta}^{(i)}, f^{(i)} = f(\boldsymbol{\uptheta}^{(i)}) \rbrace$, of the function $f$ that we want to regress. We assume that $f$ is a Gaussian process with prior mean function $m(\boldsymbol{\uptheta})$ and covariance function $\kappa(\boldsymbol{\uptheta},\boldsymbol{\uptheta}')$ also known as the kernel \citep[see][]{RasmussenWilliams2006}. The joint probability distribution of the training set is therefore $\p(\textbf{f}|\boldsymbol{\Theta}) \propto \exp\left[ \ell(\textbf{f}|\boldsymbol{\Theta}) \right]$, where the exponent $\ell(\textbf{f}|\boldsymbol{\Theta})$ is
\begin{equation}
- \frac{1}{2} \sum_{i,j=1}^t \left[f^{(i)}-m(\boldsymbol{\uptheta}^{(i)})\right]^\intercal \kappa(\boldsymbol{\uptheta}^{(i)},\boldsymbol{\uptheta}^{(j)})^{-1} \left[f^{(j)}-m(\boldsymbol{\uptheta}^{(j)})\right] .
\end{equation}
The mean function $m(\boldsymbol{\uptheta})$ and the kernel $\kappa(\boldsymbol{\uptheta},\boldsymbol{\uptheta}')$ define the functional shape and smoothness allowed for the prediction. Standard choices are respectively a constant and a squared exponential (the radial basis function, RBF), subject to additive Gaussian observation noise with variance $\sigma_\mathrm{n}^2$. Explicitly, $m(\boldsymbol{\uptheta}) \equiv C$ and 
\begin{equation}
\kappa(\boldsymbol{\uptheta},\boldsymbol{\uptheta}') \equiv \sigma_f^2 \exp\left[ -\frac{1}{2} \sum_p \left( \frac{\theta_p - \theta_p'}{\lambda_p} \right)^2 \right] + \sigma_\mathrm{n}^2 \, \updelta_\mathrm{K}(\boldsymbol{\uptheta},\boldsymbol{\uptheta}').
\end{equation}
The $\theta_p$ and $\theta_p'$ are the components of $\boldsymbol{\uptheta}$ and $\boldsymbol{\uptheta}'$, respectively. In the last term, $\updelta_\mathrm{K}(\boldsymbol{\uptheta},\boldsymbol{\uptheta}')$ is one if and only if $\boldsymbol{\uptheta} = \boldsymbol{\uptheta}'$ and zero otherwise. The hyperparameters are $C$, the $\lambda_p$ (the length scales controlling the amount of correlation between points, and hence the allowed wiggliness of $f$), $\sigma_f^2$ (the signal variance, i.e. the marginal variance of $f$ at a point $\boldsymbol{\uptheta}$ if the observation noise was zero), and $\sigma_\mathrm{n}^2$ (the observation noise). For the results of this paper, GP hyperparameters were learned from the training set using L-BFGS \citep{L-BFGS}, a popular optimiser for machine learning, and updated every time the training set was augmented with ten samples.

The predicted value $f_\star$ at a new point $\boldsymbol{\uptheta}_\star$ can be obtained from the fact that $(\lbrace \boldsymbol{\Theta}, \boldsymbol{\uptheta}_\star \rbrace, \lbrace \textbf{f} , f_\star \rbrace)$ form jointly a random realisation of the Gaussian process $f$. Thus, the target pdf $\p(f_\star|\textbf{f}, \boldsymbol{\Theta}, \boldsymbol{\uptheta}_\star)$ can be obtained from conditioning the joint pdf $\p(\textbf{f},f_\star | \boldsymbol{\Theta}, \boldsymbol{\uptheta}_\star)$ to the values of the training set $\textbf{f}$. The result is \citep[see][section 2.7]{RasmussenWilliams2006}
\begin{eqnarray}
\p(f_\star|\textbf{f}, \boldsymbol{\Theta}, \boldsymbol{\uptheta}_\star) & \propto & \exp\left[ -\frac{1}{2} \left( \frac{f_\star - \mu(\boldsymbol{\uptheta}_\star)}{\sigma(\boldsymbol{\uptheta}_\star)} \right)^2 \right], \label{eq:GP_posterior_predictive_distribution}\\
\mu(\boldsymbol{\uptheta}_\star) & \equiv & m(\boldsymbol{\uptheta}_\star) + \uline{\textbf{K}}_\star^\intercal \uuline{\textbf{K}}^{-1} (\textbf{f} - \textbf{m}), \label{eq:GP_mean}\\
\sigma^2(\boldsymbol{\uptheta}_\star) & \equiv & K_{\star\star} - \uline{\textbf{K}}_\star^\intercal \uuline{\textbf{K}}^{-1} \uline{\textbf{K}}_\star, \label{eq:GP_variance}
\end{eqnarray}
where we use the definitions
\begin{eqnarray}
K_{\star\star} & \equiv & \kappa(\boldsymbol{\uptheta}_\star, \boldsymbol{\uptheta}_\star), \label{eq:GP_notation_def_1}\\
\textbf{m} & \equiv & \left(m(\boldsymbol{\uptheta}^{(i)})\right)^\intercal \quad \mathrm{for}~\boldsymbol{\uptheta}^{(i)} \in \boldsymbol{\Theta}, \label{eq:GP_notation_def_2}\\
\uline{\textbf{K}}_\star & \equiv & \left(\kappa(\boldsymbol{\uptheta}_\star, \boldsymbol{\uptheta}^{(i)})\right)^\intercal \quad \mathrm{for}~\boldsymbol{\uptheta}^{(i)} \in \boldsymbol{\Theta}, \label{eq:GP_notation_def_3}\\
(\uuline{\textbf{K}})_{ij} & \equiv & \kappa(\boldsymbol{\uptheta}^{(i)}, \boldsymbol{\uptheta}^{(j)}) \quad \mathrm{for}~\lbrace \boldsymbol{\uptheta}^{(i)}, \boldsymbol{\uptheta}^{(j)} \rbrace \in \boldsymbol{\Theta}^2. \label{eq:GP_notation_def_4}
\end{eqnarray}

\subsection{Bayesian optimisation}
\label{ssec:Bayesian optimisation}

\begin{figure*}
\begin{center}
\includegraphics[width=0.49\textwidth]{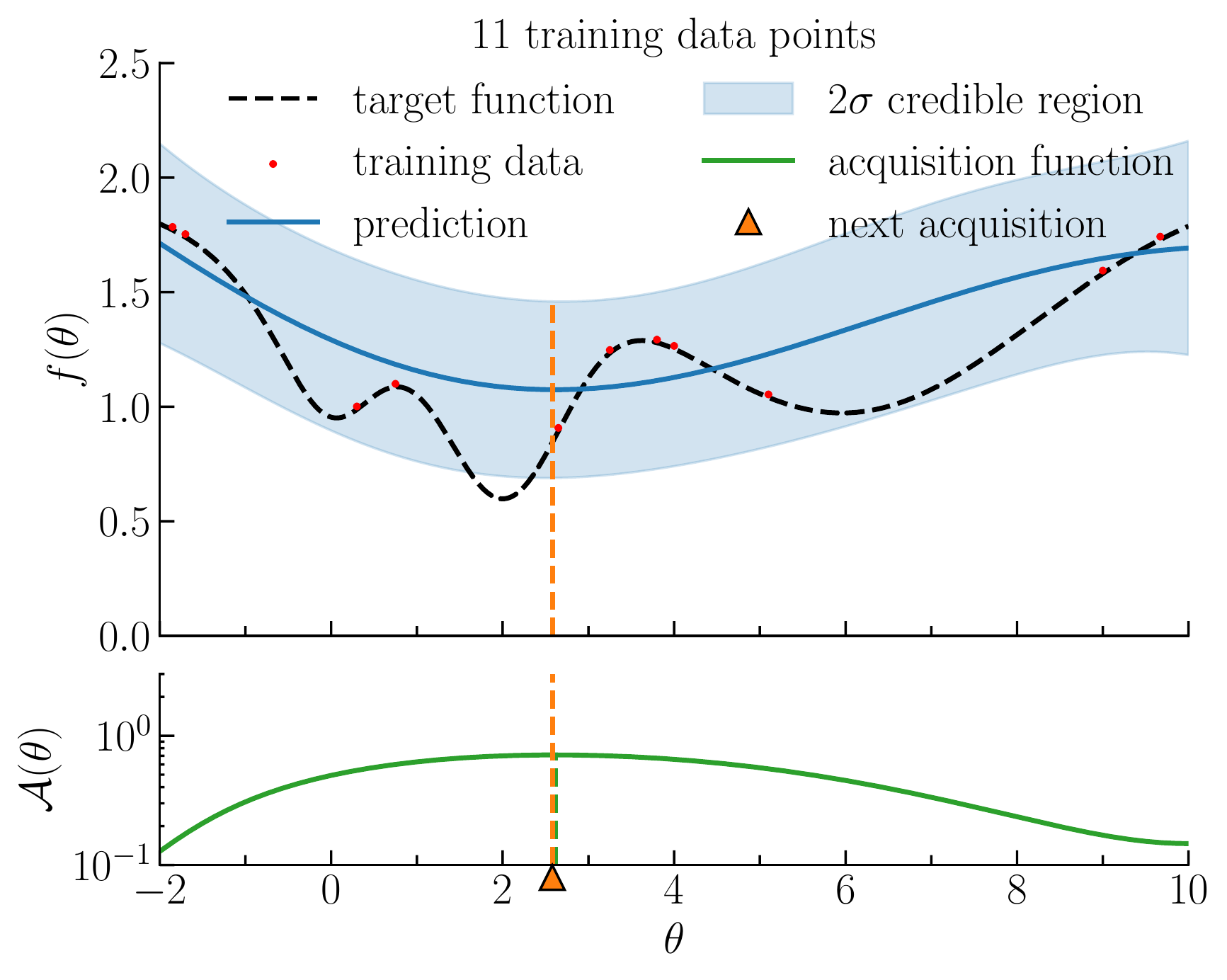} 
\includegraphics[width=0.49\textwidth]{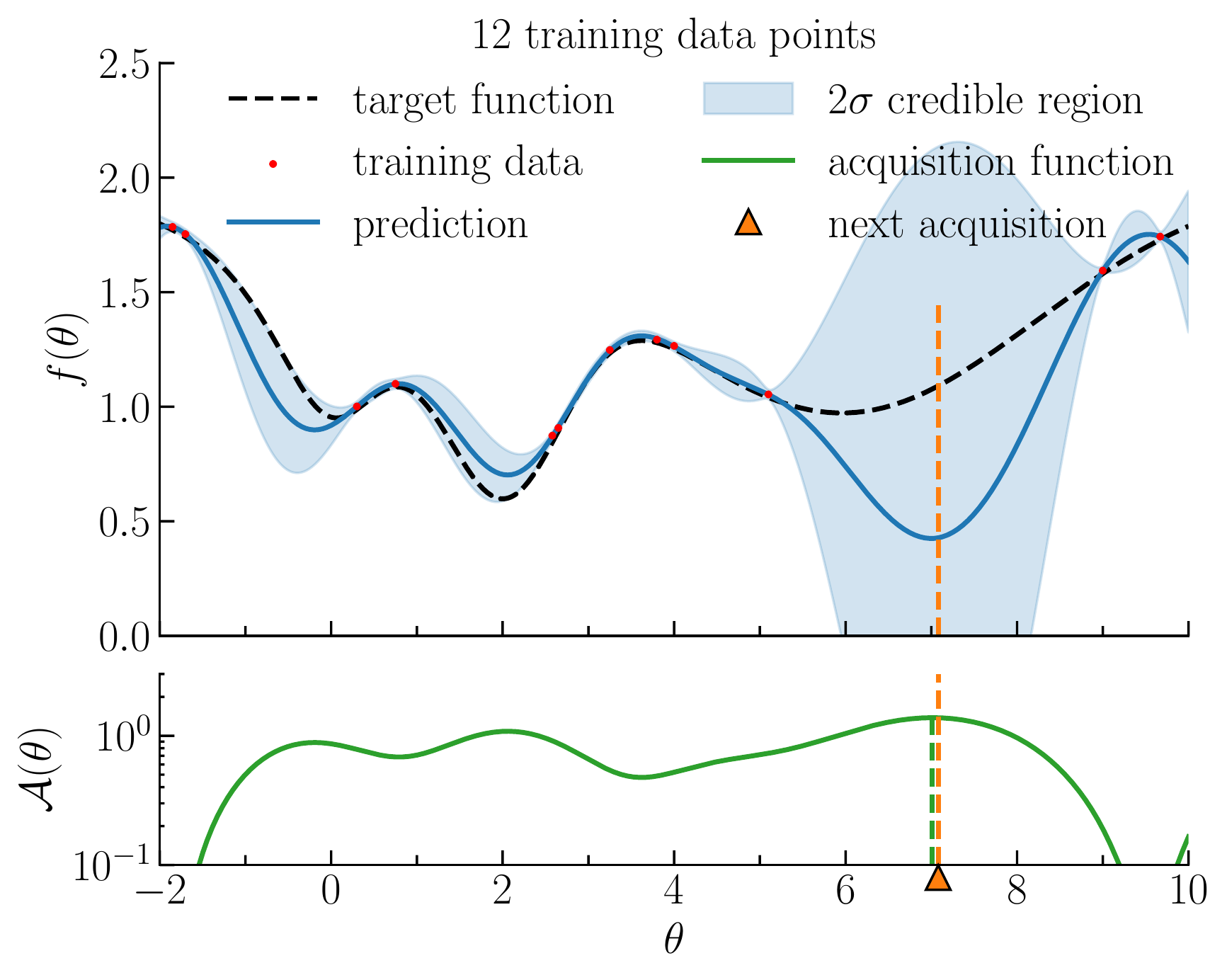} \\
\includegraphics[width=0.49\textwidth]{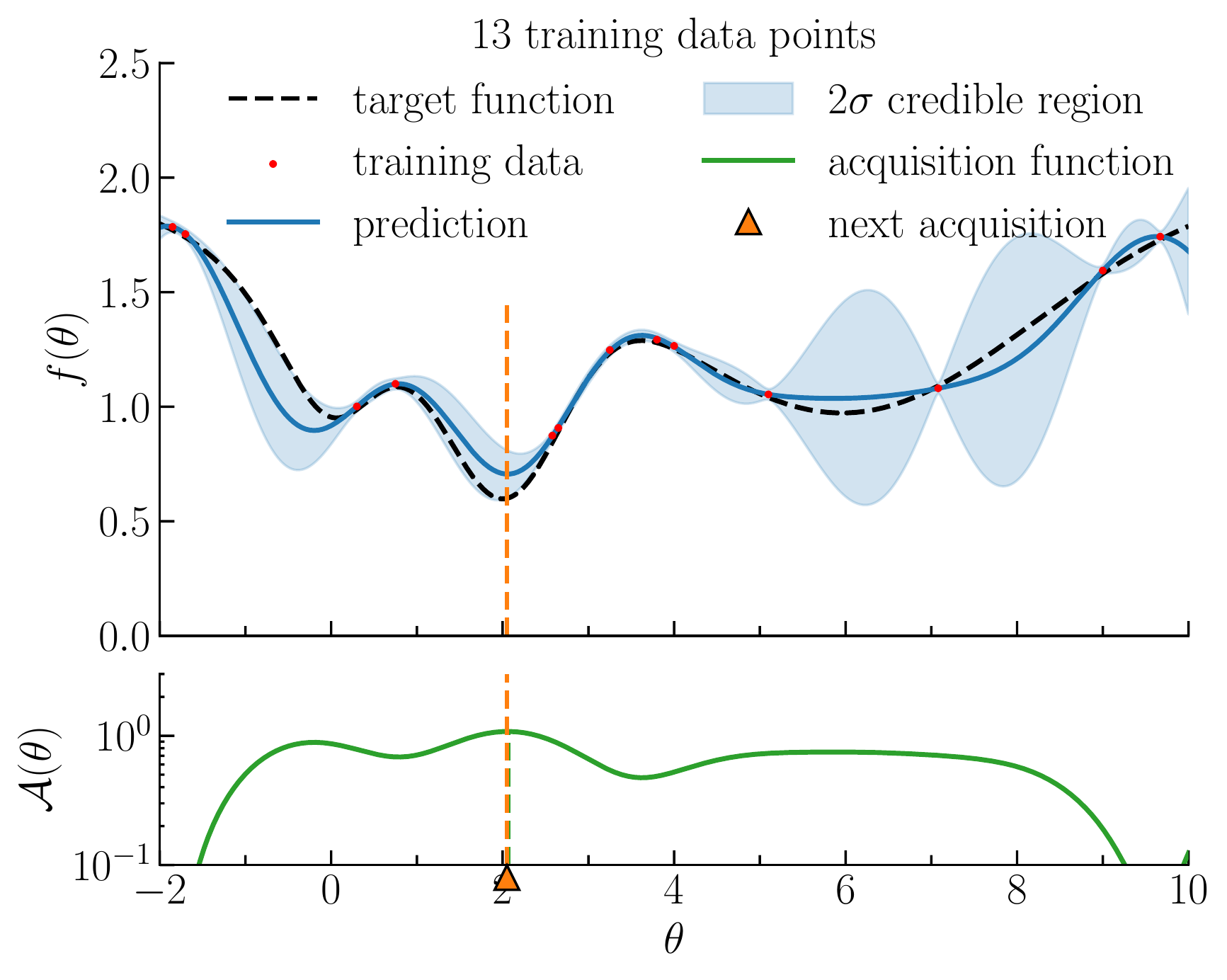} 
\includegraphics[width=0.49\textwidth]{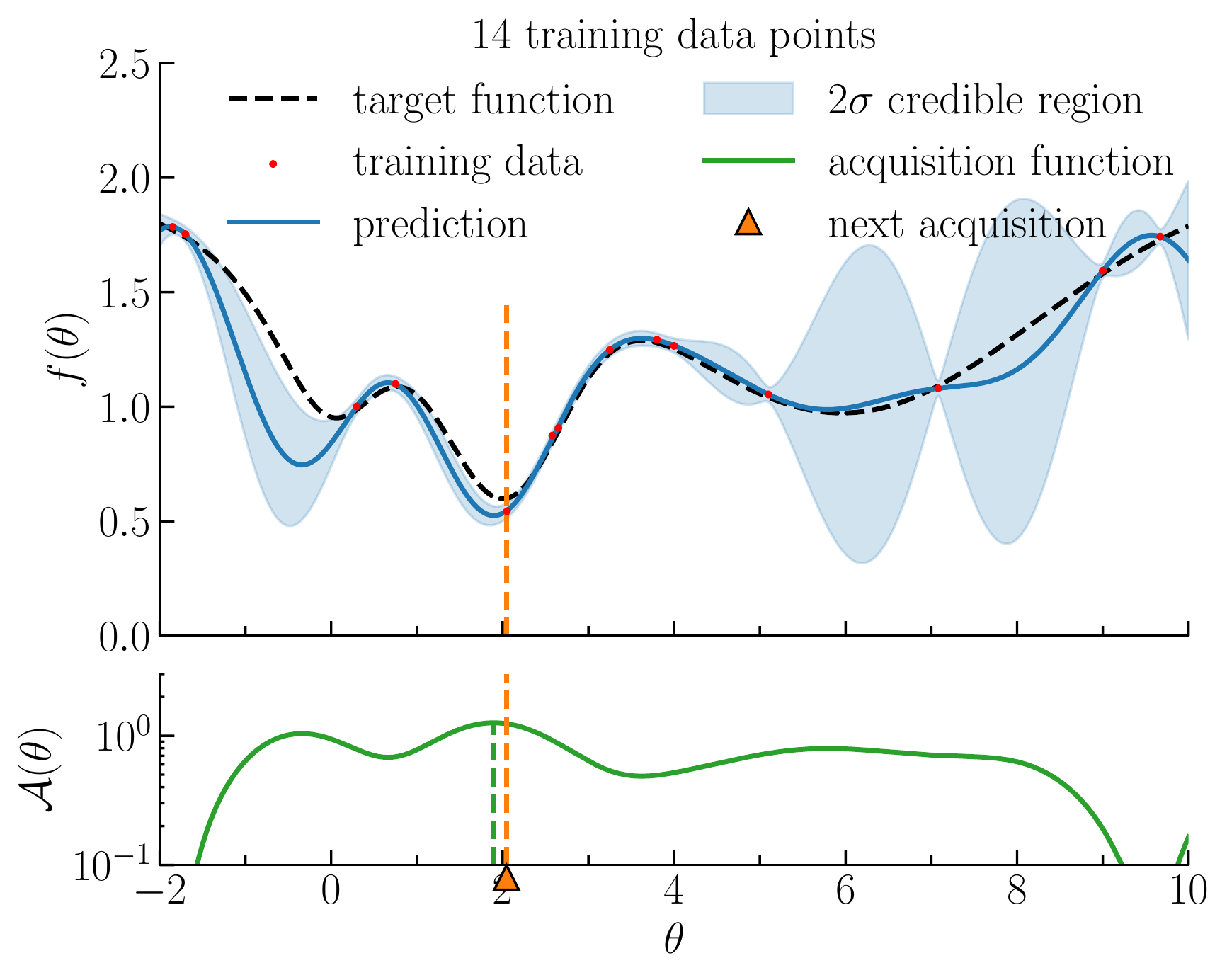} 
\caption{Illustration of four consecutive steps of Bayesian optimisation to learn the test function of figure \ref{fig:GP_illustration}. For each step, the top panel shows the training data points (red dots) and the regression (blue line and shaded region). The bottom panel shows the acquisition function (the expected improvement, solid green line) with its maximiser (dashed green line). The next acquisition point, i.e. where to run a simulation to be added to the training set, is shown in orange; it differs from the maximiser of the acquisition function by a small random number. The acquisition function used is the expected improvement, aiming at finding the minimum of $f$. Hyperparameters of the regression kernel are optimised after each acquisition. As can observed, Bayesian optimisation implements a trade-off between exploration (evaluation of the target function where the variance is large, e.g. after 12 points) and exploitation (evaluation of the target function close to the predicted minimum, e.g. after 11, 13, and 14 points).  \label{fig:BO_illustration}}
\end{center}
\end{figure*}

The second major ingredient of the proposed approach is Bayesian optimisation, which allows the inference of the regression function $J(\boldsymbol{\uptheta})$ while avoiding unnecessary computations. It allows active construction of the training data set $\lbrace (\boldsymbol{\uptheta}^{(i)} , \Delta_{\boldsymbol{\uptheta}}^{(i)}) \rbrace$, updating the proposal of new points using the regressed $\widehat{J}^{(\mathrm{t})}(\boldsymbol{\uptheta})$ (thus addressing issue 3 with likelihood-free rejection sampling, see section \ref{ssec:Computational difficulties with likelihood-free rejection sampling}). Further, since we are mostly interested in the regions of the parameter space where the variance of the approximate posterior is large (due to its stochasticity), the acquisition rules can prioritise these regions, so as to obtain a better approximation of $J(\boldsymbol{\uptheta})$ there (thus addressing issue 4).

Bayesian optimisation is a decision-making framework under uncertainty, for the automatic learning of unknown functions. It aims at gathering training data in such a manner as to evaluate the regression model the least number of times while revealing as much information as possible about the target function and, in particular, the location of the optimum or optima. The method proceeds by iteratively picking predictors to be probed (i.e. simulations to be run) in a manner that trades off \textit{exploration} (parameters for which the outcome is most uncertain) and \textit{exploitation} (parameters which are expected to have a good outcome for the targeted application). In many contexts, Bayesian optimisation has been shown to obtain better results with fewer simulations than grid search or random search, due to its ability to reason about the interest of simulations before they are run \citep[see][for a review]{Brochu2010}. Figure \ref{fig:BO_illustration} illustrates Bayesian optimisation in combination with Gaussian process regression, applied to finding the minimum of the test function of figure \ref{fig:GP_illustration}.

In the following, we give a brief overview of the elements of Bayesian optimisation used in this paper. In order to add a new point to the training data set $(\boldsymbol{\Theta}, \textbf{f}) \equiv \lbrace (\boldsymbol{\uptheta}^{(i)}, f^{(i)} = f(\boldsymbol{\uptheta}^{(i)}) \rbrace$, Bayesian optimisation uses an acquisition function $\mathcal{A}(\boldsymbol{\uptheta})$ that estimates how useful the evaluation of the simulator at $\boldsymbol{\uptheta}$ will be in order to learn the target function. The acquisition function is constructed from the posterior predictive distribution of $f$ given the training set $(\boldsymbol{\Theta}, \textbf{f})$, i.e. from the mean prediction $\mu(\boldsymbol{\uptheta})$ and the uncertainty $\sigma(\boldsymbol{\uptheta})$ of the regression analysis (equations \eqref{eq:GP_mean} and \eqref{eq:GP_variance}). The optimum of the acquisition function in parameter space determines the next point $\boldsymbol{\uptheta}_\star \equiv \mathrm{argopt}_{\boldsymbol{\uptheta}} \mathcal{A}(\boldsymbol{\uptheta})$ to be evaluated by the simulator ($\mathrm{argopt} = \mathrm{argmax}$ or $\mathrm{argmin}$ depending on how the acquisition function is defined), so that the training set can be augmented with $(\boldsymbol{\uptheta}_\star, f(\boldsymbol{\uptheta}_\star))$. The acquisition function is a scalar function whose evaluation should be reasonably expensive, so that its optimum can be found by simple search methods such as gradient descent. 

The algorithm needs to be initialised with an initial training set. In numerical experiments, we found that building this initial set by drawing from the prior (as would typically be done in likelihood-free rejection sampling) can result in difficulties with the first iterations of Gaussian process regression. Uniformly-distributed points within the boundaries of the GP are also a poor choice, as they will result in an uneven initial sampling of the parameter space. To circumvent this issue, we build the initial training set using a low-discrepancy quasi-random Sobol sequence \citep{Sobol1967}, which covers the parameter space more evenly.

\subsection{Expressions for the approximate posterior}
\label{ssec:Expressions for the approximate posterior}

As discussed in section \ref{ssec:Regression of the discrepancy}, using $\Delta_{\boldsymbol{\uptheta}}^{\textbf{C}_{\boldsymbol{\uptheta}}}$ as the regressed quantity directly gives an estimate of $J(\boldsymbol{\uptheta})$ in equation \eqref{eq:def_J}. The response variable is thus $f(\boldsymbol{\uptheta}) \equiv \Delta_{\boldsymbol{\uptheta}}^{\textbf{C}_{\boldsymbol{\uptheta}}}$ and the regression then gives
\begin{equation}
\widehat{J}^{(\mathrm{t})}(\boldsymbol{\uptheta}) = \mu(\boldsymbol{\uptheta}).
\label{eq:J_t_equals_mu}
\end{equation}

In the parametric approach to likelihood approximation, this is equivalent to an approximation of $-2\tilde{\ell}(\boldsymbol{\uptheta}) = -2\log \widetilde{L}(\boldsymbol{\uptheta})$ (see equation \eqref{eq:l_J_proposition_1}). The expectation of the (unnormalised) approximate posterior is therefore directly given as (see equation \eqref{eq:approx_problem_Bayes})
\begin{equation}
\mathrm{E}^{(\mathrm{t})} \left[ \p_{\bolfi}(\boldsymbol{\uptheta}|\boldsymbol{\Phi}_\mathrm{O}, \textbf{f}, \boldsymbol{\Theta}) \right] \equiv \p(\boldsymbol{\uptheta}) \exp\left( -\frac{1}{2} \mu(\boldsymbol{\uptheta}) \right),
\label{eq:approximate_posterior_expectation}
\end{equation}
where $\p_{\bolfi}(\boldsymbol{\uptheta}|\boldsymbol{\Phi}_\mathrm{O}, \textbf{f}, \boldsymbol{\Theta}) \approx Z_{\boldsymbol{\Phi}} \times \p(\boldsymbol{\uptheta}|\boldsymbol{\Phi})_{|\boldsymbol{\Phi}=\boldsymbol{\Phi}_\mathrm{O}}$.

The estimate of the variance of $f(\boldsymbol{\uptheta})$ can also be propagated to the approximate posterior, giving
\begin{equation}
\mathrm{V}^{(\mathrm{t})} \left[ \p_{\bolfi}(\boldsymbol{\uptheta}|\boldsymbol{\Phi}_\mathrm{O}, \textbf{f}, \boldsymbol{\Theta}) \right] \equiv \frac{\p(\boldsymbol{\uptheta})^2}{4} \exp\left[ -\mu(\boldsymbol{\uptheta}) \right] \sigma^2(\boldsymbol{\uptheta}) .
\label{eq:approximate_posterior_variance}
\end{equation}
Details of the computations can be found in appendix \ref{sapx:Expressions for the approximate posterior}.

Expressions for the {\bolfi} posterior in the non-parametric approach with the uniform kernel can also be derived \citep[][lemma 3.1]{Jaervenpaeae2017}. As this paper focuses on the parametric approach, we refer to the literature for the former case.

\subsection{Acquisition rules}
\label{ssec:Acquisition rules}

\subsubsection{Expected improvement}
\label{sssec:Expected improvement}

Standard Bayesian optimisation uses acquisition functions that estimate how useful the next evaluation of the simulator will be in order to find the minimum or minima of the target function. While several other choices are possible \citep[see e.g.][]{Brochu2010}, in this work we discuss the acquisition function known as \textit{expected improvement} (EI). The \textit{improvement} is defined by $I(\boldsymbol{\uptheta}_\star) = \max\left[\min(\textbf{f}) - f(\boldsymbol{\uptheta}_\star), 0\right]$, and the expected improvement is $\mathrm{EI}(\boldsymbol{\uptheta}_\star) \equiv \mathrm{E}^{(\mathrm{t})}\left[ I(\boldsymbol{\uptheta}_\star) \right]$, where the expectation is taken with respect to the random observation assuming decision $\boldsymbol{\uptheta}_\star$. For a Gaussian process regressor, this evaluates to \citep[see][section 2.3]{Brochu2010}
\begin{equation}
\mathrm{EI}(\boldsymbol{\uptheta}_\star) \equiv \sigma(\boldsymbol{\uptheta}_\star) \left[ z\Phi(z) + \phi(z) \right], \, \mathrm{with}~z \equiv \frac{\min(\textbf{f}) - \mu(\boldsymbol{\uptheta}_\star)}{\sigma(\boldsymbol{\uptheta}_\star)},
\label{eq:EI}
\end{equation}
or $\mathrm{EI}(\boldsymbol{\uptheta}_\star) \equiv 0$ if $\sigma(\boldsymbol{\uptheta}_\star)=0$, where $\phi$ and $\Phi$ denote respectively the pdf and the cumulative distribution function (cdf) of the unit-variance zero-mean Gaussian. The decision rule is to select the location $\boldsymbol{\uptheta}_\star$ that maximises $\mathrm{EI}(\boldsymbol{\uptheta}_\star)$.

The EI criterion can be interpreted as follows: since the goal is to find the minimum of $f$, a reward equal to the improvement $\min(\textbf{f}) - f(\boldsymbol{\uptheta}_\star)$ is received if $f(\boldsymbol{\uptheta}_\star)$ is smaller than all the values observed so far, otherwise no reward is received. The first term appearing in equation \eqref{eq:EI} is maximised when evaluating at points with high uncertainty (exploration); and, at fixed variance, the second term is maximised by evaluating at points with low mean (exploitation). The expected improvement therefore automatically captures the exploration-exploitation trade-off as a result of the Bayesian decision-theoretic treatment.

\subsubsection{Expected integrated variance}
\label{sssec:Expected integrated variance}

As pointed out by \citet{Jaervenpaeae2017}, in Bayesian optimisation for approximate Bayesian computation, the goal should not be to find the minimum of $J(\boldsymbol{\uptheta})$, but rather to minimise the expected uncertainty in the estimate of the approximate posterior over the future evaluation of the simulator at $\boldsymbol{\uptheta}_\star$. Consequently, they propose an acquisition function, known as the \textit{expected integrated variance} (ExpIntVar or EIV in the following) that selects the next evaluation location to minimise the expected variance of the future posterior density $\p_{\bolfi}(\boldsymbol{\uptheta}|\boldsymbol{\Phi}_\mathrm{O}, \textbf{f}, \boldsymbol{\Theta}, \boldsymbol{\uptheta}_\star)$ over the parameter space. The framework used is Bayesian decision theory. Formally, the loss due to our uncertain knowledge of the approximate posterior density can be defined as
\begin{equation}
\mathpzc{L}\left[ \p_{\bolfi}(\boldsymbol{\uptheta}|\boldsymbol{\Phi}_\mathrm{O}, \textbf{f}, \boldsymbol{\Theta}) \right] = \int \mathrm{V}^{(\mathrm{t})}\left[ \p_{\bolfi}(\boldsymbol{\uptheta}|\boldsymbol{\Phi}_\mathrm{O}, \textbf{f}, \boldsymbol{\Theta}) \right] \, \mathrm{d}\boldsymbol{\uptheta},
\end{equation}
and the acquisition rule is to select the location $\boldsymbol{\uptheta}_\star$ that minimises
\begin{equation}
\begin{split}
& \mathrm{EIV}(\boldsymbol{\uptheta}_\star) \equiv \mathrm{E}^{(\mathrm{t})} \left[ \mathpzc{L}\left[ \p_{\bolfi}(\boldsymbol{\uptheta}|\boldsymbol{\Phi}_\mathrm{O}, \textbf{f}, \boldsymbol{\Theta}, f_\star, \boldsymbol{\uptheta}_\star) \right] \right] \\
& = \int \mathpzc{L}\left[ \p_{\bolfi}(\boldsymbol{\uptheta}|\boldsymbol{\Phi}_\mathrm{O}, \textbf{f}, \boldsymbol{\Theta}, f_\star, \boldsymbol{\uptheta}_\star) \right] \p(f_\star|\textbf{f}, \boldsymbol{\Theta}, \boldsymbol{\uptheta}_\star) \,  \mathrm{d}f_\star
\end{split}
\end{equation}
with respect to $\boldsymbol{\uptheta}_\star$, where we have to marginalise over the unknown simulator output $f_\star$ using the probabilistic model $\p(f_\star|\textbf{f}, \boldsymbol{\Theta}, \boldsymbol{\uptheta}_\star)$ (equations \eqref{eq:GP_posterior_predictive_distribution}--\eqref{eq:GP_variance}).

\citet[][proposition 3.2]{Jaervenpaeae2017} derive the expressions for the expected integrated variance for a GP model in the non-parametric approach. In appendix \ref{apx:Derivations of the mathematical results}, we extend this work and derive the ExpIntVar acquisition function and its gradient in the parametric approach. The result is the following: under the GP model, the expected integrated variance after running the simulation model with parameter $\boldsymbol{\uptheta}_\star$ is given by
\begin{equation}
\mathrm{EIV}(\boldsymbol{\uptheta}_\star) = \int \frac{\p(\boldsymbol{\uptheta})^2}{4} \exp\left[ -\mu(\boldsymbol{\uptheta}) \right] \left[ \sigma^2(\boldsymbol{\uptheta}) - \tau^2(\boldsymbol{\uptheta},\boldsymbol{\uptheta}_\star) \right] \, \mathrm{d}\boldsymbol{\uptheta},
\label{eq:EIV}
\end{equation}
with
\begin{equation}
\tau^2(\boldsymbol{\uptheta},\boldsymbol{\uptheta}_\star) \equiv \dfrac{\mathrm{cov}^2(\boldsymbol{\uptheta},\boldsymbol{\uptheta}_\star)}{\sigma^2(\boldsymbol{\uptheta}_\star)},
\label{eq:def_tau}
\end{equation}
where $\mathrm{cov}(\boldsymbol{\uptheta},\boldsymbol{\uptheta}_\star) \equiv \kappa(\boldsymbol{\uptheta},\boldsymbol{\uptheta}_\star) - \uline{\textbf{K}}^\intercal \uuline{\textbf{K}}^{-1}\vspace{-4pt} \uline{\textbf{K}}_\star$ is the GP posterior predicted covariance between the evaluation point $\boldsymbol{\uptheta}$ in the integral and the candidate location for the next evaluation $\boldsymbol{\uptheta}_\star$. Note that in addition to the notations given by equations \eqref{eq:GP_notation_def_1}--\eqref{eq:GP_notation_def_4}, we have introduced the vector
\begin{equation}
\uline{\textbf{K}} \equiv \left(\kappa(\boldsymbol{\uptheta}, \boldsymbol{\uptheta}^{(i)})\right)^\intercal \quad \mathrm{for}~\boldsymbol{\uptheta}^{(i)} \in \boldsymbol{\Theta}.
\label{eq:GP_notation_def_5}
\end{equation}

It is of interest to examine when the integrand in equation \eqref{eq:EIV} is small. As for the EI (equation \eqref{eq:EI}), optimal values are found when the mean of the discrepancy $\mu(\boldsymbol{\uptheta})$ is small or the variance $\sigma^2(\boldsymbol{\uptheta})$ is large. This effect is what yields the trade-off between exploitation and exploration for the ExpIntVar acquisition rule. However, unlike in standard Bayesian optimisation strategies such as the EI, the trade-off is a non-local process (due to the integration over the parameter space), and also depends on the prior, so as to minimise the uncertainty in the posterior (and not likelihood) approximation.

Computing the expected integrated variance requires integration over the parameter space. In this work, the integration is performed on a regular grid of $50$ points per dimension within the GP boundaries. In high dimension, the integral can become prohibitively expensive to compute on a grid. As discussed by \citet{Jaervenpaeae2017}, it can then be evaluated with Monte Carlo or quasi-Monte Carlo methods such as importance sampling.

In numerical experiments, we have found that the ExpIntVar criterion (as any acquisition function for Bayesian optimisation) has some sensitivity to the initial training set. In particular, the initial set (built from a Sobol sequence or otherwise) shall sample sufficiently well the GP domain, which shall encompass the prior. This ensures that the prior volume is never wider than the training data. Under this condition, as \citet{Jaervenpaeae2017}, we have found that ExpIntVar is stable, in the sense that it produces consistent {\bolfi} posteriors over different realisations of the initial training data set and simulator outputs.

\subsubsection{Stochastic versus deterministic acquisition rules}
\label{sssec:Stochastic versus deterministic acquisition rules}

The above rules do not guarantee that the selected $\boldsymbol{\uptheta}_\star$ is different from a previously acquired $\boldsymbol{\uptheta}^{(i)}$. \citet[][see in particular appendix C]{GutmannCorander2016} found that this can result in a poor exploration of the parameter space, and propose to add a stochastic element to the decision rule in order to avoid getting stuck at one point. In some experiments, we followed this prescription by adding an ``acquisition noise'' of strength $\sigma_\mathrm{a}^p$ to each component of the optimiser of the acquisition function. More precisely, $\boldsymbol{\uptheta}_\star$ is sampled from the Gaussian distribution $\mathpzc{G}(\boldsymbol{\uptheta}_\mathrm{opt}, \textbf{D})$, where $\boldsymbol{\uptheta}_\mathrm{opt} \equiv \mathrm{argopt}_{\boldsymbol{\uptheta}} \mathcal{A}(\boldsymbol{\uptheta})$ and $\textbf{D}$ is the diagonal covariance matrix of components $(\sigma_\mathrm{a}^p)^2$. The $\sigma_\mathrm{a}^p$ are chosen to be of order $\lambda_p/10$.

For a more extensive discussion and comparison of various stochastic and deterministic acquisition rules, the reader is referred to \citet{Jaervenpaeae2017}.

\section{Applications}
\label{sec:Applications}

In this section, we show the application of {\bolfi} to several application studies. In particular, we discuss the simulator and the computable approximation of the likelihood to be used, and compare {\bolfi} to likelihood-free rejection sampling in terms of computational efficiency. In all cases, we show that {\bolfi} reduces the amount of required simulations by several orders of magnitude.

In section \ref{ssec:Summarising Gaussian signals}, we discuss the toy problem of summarising Gaussian signals (i.e. inferring the unknown mean and/or variance of Gaussian-distributed data). In section \ref{ssec:Supernova cosmology}, we show the first application of {\bolfi} to a real cosmological problem using actual observational data: the inference of cosmological parameters from supernovae data. For each test case, we refer to the corresponding section in the appendices for the details of the data model and inference assumptions.

\subsection{Summarising Gaussian signals}
\label{ssec:Summarising Gaussian signals}

A simple toy model can be constructed from the general problem of summarising Gaussian signals with unknown mean, or with unknown mean and variance. This example allows for the comparison of {\bolfi} and likelihood-free rejection sampling to the true posterior conditional on the full data, which is known analytically. All the details of this model are given in appendix \ref{apx:Summarising Gaussian signals}.

\subsubsection{Unknown mean, known variance}
\label{sssec:Unknown mean, known variance}

\begin{figure}
\begin{center}
\includegraphics[width=\columnwidth]{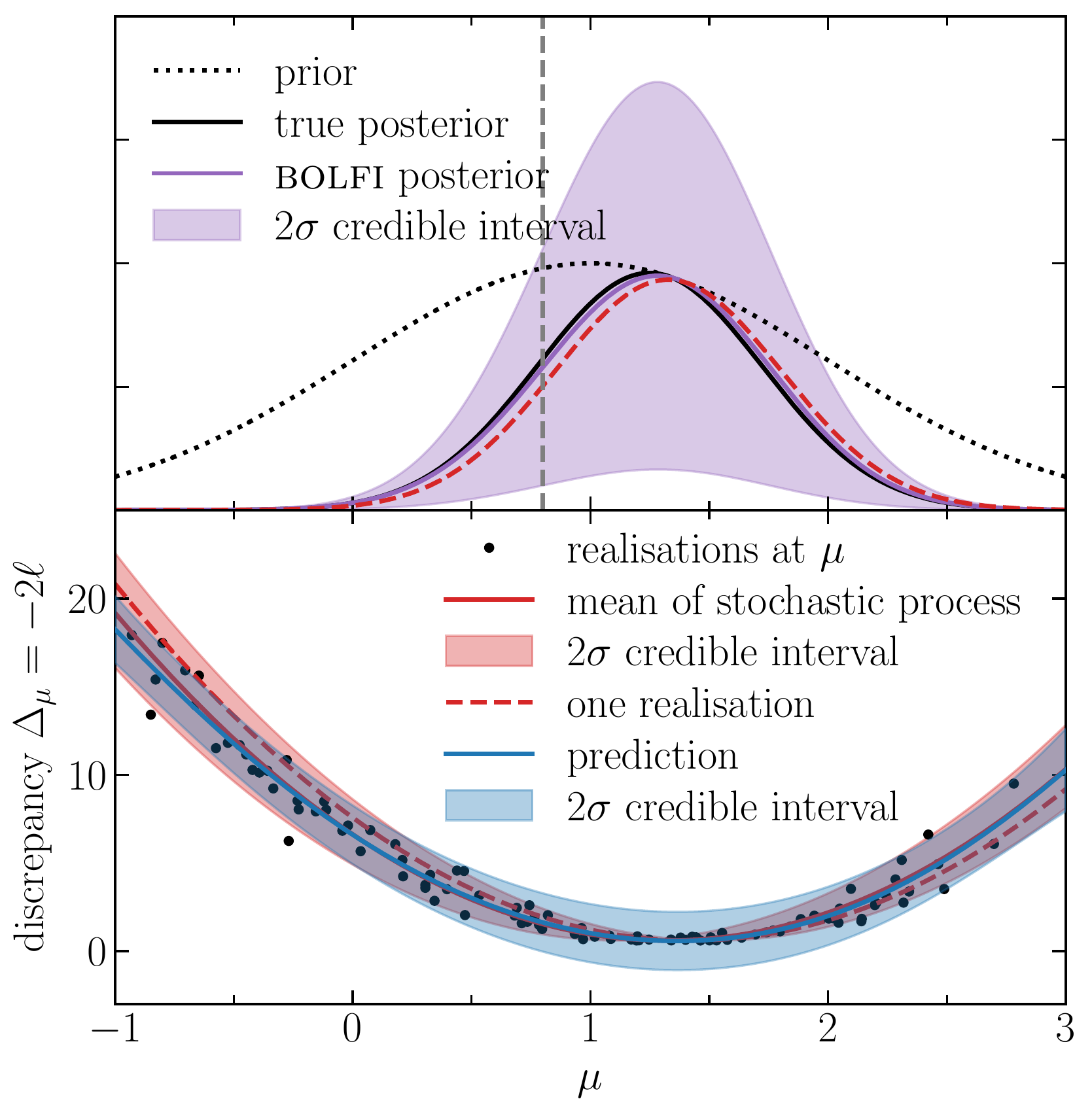} 
\caption{Illustration of {\bolfi} for a one-dimensional problem, the inference of the unknown mean $\mu$ of a Gaussian. \textit{Lower panel}. The discrepancy $\Delta_\mu$ (i.e. twice the negative log-likelihood) is a stochastic process due to the limited computational resources. Its mean and the $2\sigma$ credible interval are shown in red. The dashed red line shows one realisation of the stochastic process as a function of $\mu$. Simulations at different $\mu$ are shown as black dots. {\bolfi} builds a probabilistic model for the discrepancy, the mean and $2\sigma$ credible interval of which are shown in blue. \textit{Upper panel}. The expectation of the (rescaled) {\bolfi} posterior and its $2\sigma$ credible interval are shown in comparison to the exact posterior for the problem. The dashed red line shows the posterior obtained from the corresponding realisation of the stochastic process of the lower panel. \label{fig:Gaussian_mean_illustration}}
\end{center}
\end{figure}

We first consider the problem, already discussed by \citet{GutmannCorander2016}, where the data $\textbf{d}$ are a vector of $n$ components drawn from a Gaussian with unknown mean $\mu$ and known variance $\sigma^2_\mathrm{true}$. The empirical mean $\Phi^1$ is a sufficient summary statistic for the problem of inferring $\mu$. The distribution of simulated $\Phi^1_\mu$ takes a simple form, $\Phi^1_\mu \sim \mathpzc{G}\left( \mu, \sigma^2_\mathrm{true}/n \right)$. Using here the true variance, the discrepancy and synthetic likelihood are
\begin{equation}
\Delta^1_\mu = -2 \hat{\ell}^N_1(\mu) = \log \left(\frac{2\pi \sigma^2_\mathrm{true}}{n} \right) + n\frac{(\Phi^1_\mathrm{O}-\hat{\mu}^1_\mu)^2}{\sigma^2_\mathrm{true}},
\end{equation}
where $\hat{\mu}^1_\mu$ is an average of $N$ realisations of $\Phi^1_\mu$. In figure \ref{fig:Gaussian_mean_illustration} (lower panel), the black dots show simulations of $\Delta^1_\mu$ for different values of $\mu$. We have $\hat{\mu}^1_\mu \sim \mathpzc{G}\left( \mu, \sigma^2_\mathrm{true}/(Nn) \right)$, therefore the stochastic process defining the discrepancy can be written
\begin{equation}
\Delta^1_\mu = \log \left(\frac{2\pi \sigma^2_\mathrm{true}}{n} \right) + n\frac{(\Phi^1_\mathrm{O}- \mu -g )^2}{\sigma^2_\mathrm{true}}, \quad g \sim \mathpzc{G}\left(0, \sigma^2_g \right),
\end{equation}
where $\sigma^2_g \equiv \sigma^2_\mathrm{true}/(Nn)$. Each realisation of $g$ gives a different mapping $\mu \mapsto \Delta^1_\mu$. In figure \ref{fig:Gaussian_mean_illustration}, we show one such realisation in the lower panel, and the corresponding approximate posterior in the upper panel. Using the percent point function (inverse of the cdf) of the Gaussian $\mathpzc{G}\left(0, \sigma^2_g \right)$, we also show in red the mean and $2\sigma$ credible interval of the true stochastic process.

The GP regression using the simulations shown as the training set is represented in blue in the lower panel of figure \ref{fig:Gaussian_mean_illustration}. The corresponding {\bolfi} posterior and its variance, defined by equations \eqref{eq:approximate_posterior_expectation} and \eqref{eq:approximate_posterior_variance}, are shown in purple in the upper panel. The uncertainty in the estimate of the posterior (shaded purple region) is due to the limited number of available simulations (and not to the noisiness of individual training points). It is the expectation of this uncertainty under the next evaluation of the simulator which is minimised in parameter space by the ExpIntVar acquisition rule.

\subsubsection{Unknown mean and variance}
\label{sssec:Unknown mean and variance}

\begin{figure*}
\begin{center}
\includegraphics[width=\textwidth]{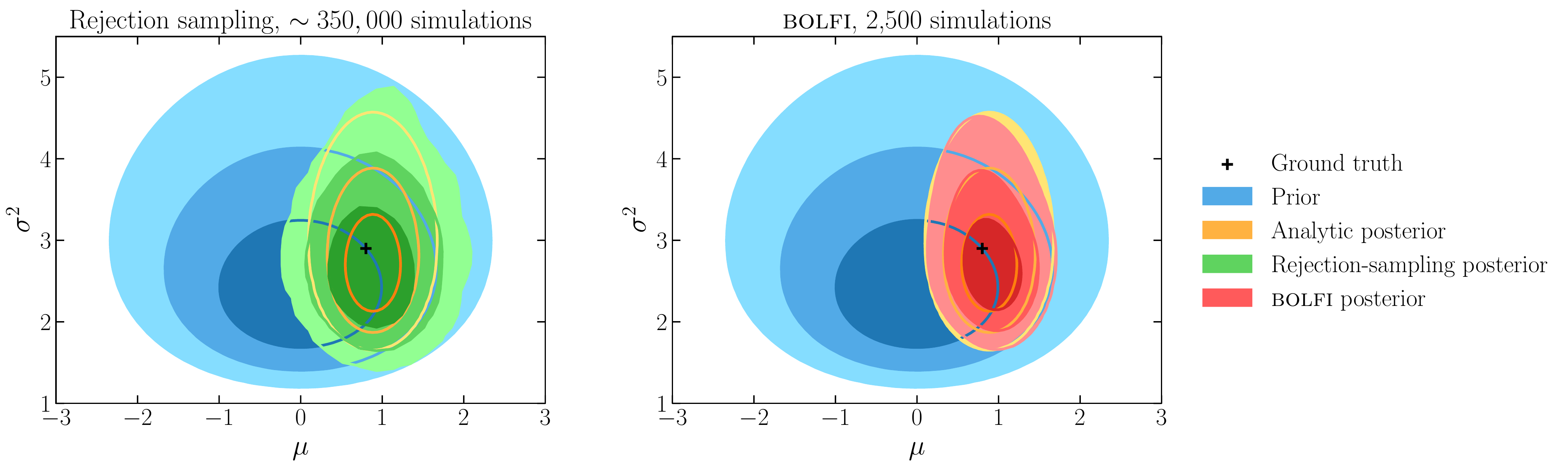} 
\caption{Prior and posterior for the joint inference of the mean and variance of Gaussian signals. The prior and exact posterior (from the analytic solution) are Gaussian-inverse-Gamma distributed and shown in blue and orange, respectively. In the left panel, the approximate rejection-sampling posterior, based on $5,000$ samples accepted out of $\sim 350,000$ simulations, is shown in green. It loosely encloses the exact posterior. In the right panel, the approximate {\bolfi} posterior, based on $2,500$ simulations only, is shown in red. It is a much finer approximation of the exact posterior. For all distributions, the $1\sigma$, $2\sigma$ and $3\sigma$ contours are shown.\label{fig:Gaussian_mean_variance}}
\end{center}
\end{figure*}

We now consider the problem where the full data set $\textbf{d}$ is a vector of $n$ components drawn from a Gaussian with unknown mean $\mu$ and unknown variance $\sigma^2$. The aim is the two-dimensional inference of $\boldsymbol{\uptheta} \equiv (\mu, \sigma^2)$. Evidently, the true likelihood $\mathcal{L}(\mu, \sigma^2)$ for this problem is the Gaussian characterised by $(\mu, \sigma^2)$. The Gaussian-inverse-Gamma distribution is the conjugate prior for this likelihood. It is described by four parameters. Adopting a Gaussian-inverse-Gamma prior characterised by $(\alpha, \beta, \eta, \lambda)$ yields a Gaussian-inverse-Gamma posterior characterised by $(\alpha', \beta', \eta', \lambda')$ given by equations \eqref{eq:Gaussian_analytic_solution_alpha}--\eqref{eq:Gaussian_analytic_solution_lambda}. This is the analytic solution to which we compare our approximate results.

For the numerical approach, we forward model the problem using a simulator that draws from the prior, simulates $N = 10$ realisations of the Gaussian signal, and compresses them to two summary statistics, the empirical mean and variance, respectively $\Phi^1$ and $\Phi^2$. The graphical probabilistic model is given in figure \ref{fig:BHM_Gaussian_model}. It is a noise-free simulator without latent variables (of the type given by figure \ref{fig:BHM_exact}, right) completed by a deterministic compression of the full data. Note that the vector $\boldsymbol{\Phi} \equiv (\Phi^1 , \Phi^2)$ is a sufficient statistic for the inference of $(\mu, \sigma^2)$. To perform likelihood-free inference, we also need a computable approximation $\widehat{L}^N(\mu, \sigma^2)$ of the true likelihood. We derive such an approximation in section \ref{sapx:Derivation of the Gaussian-Gamma synthetic likelihood for likelihood-free inference} using a parametric approach, under the assumptions (exactly verified in this example) that $\Phi^1$ is Gaussian-distributed and $\Phi^2$ is Gamma-distributed. We name it the Gaussian-Gamma synthetic likelihood.

The posterior obtained from likelihood-free rejection sampling is shown in green in figure \ref{fig:Gaussian_mean_variance} (left) in comparison to the prior (in blue) and the analytic posterior (in orange). It was obtained from $5,000$ accepted samples using a threshold of $\varepsilon = 4$ on $-2\hat{\ell}^N$. The entire run required $\sim 350,000$ forward simulations in total, the vast majority of which have been rejected. The rejection-sampling posterior is a fair approximation to the true posterior, unbiased but broader, as expected from a rejection-sampling method. 

For comparison, the posterior obtained via {\bolfi} is shown in red in figure \ref{fig:Gaussian_mean_variance} (right). {\bolfi} was initialised using a Sobol sequence of $20$ members to compute the original surrogate surface, and Bayesian optimisation with the ExpIntVar acquisition function and acquisition noise was run to acquire $230$ more samples. As can be observed, {\bolfi} allows very precise likelihood-free inference; in particular, the $1\sigma$, $2\sigma$ and $3\sigma$ contours (the latter corresponding to the $0.27\%$ least likely events) of the analytic posterior are reconstructed almost perfectly. The overall cost to get these results is only $2,500$ simulations with {\bolfi} versus $\sim 350,000$ with rejection sampling (for a poorer approximation of the analytic posterior), which corresponds to a reduction by $2$ orders of magnitude. 

\subsection{Supernova cosmology}
\label{ssec:Supernova cosmology}

\begin{figure*}
\begin{center}
\includegraphics[width=\textwidth]{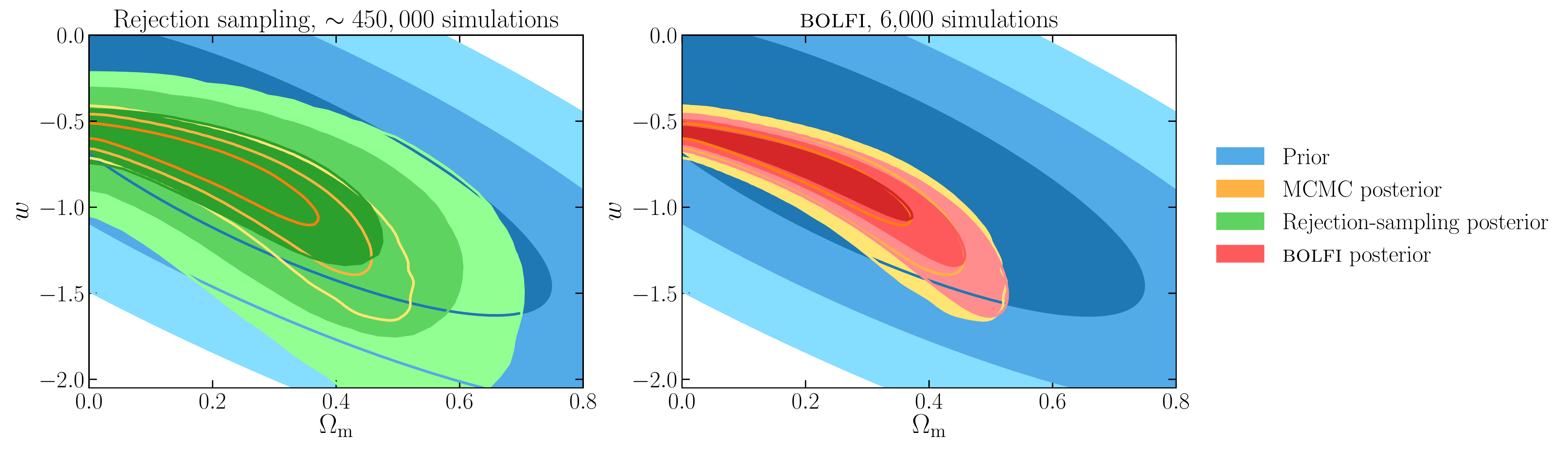} 
\caption{Prior and posterior distributions for the joint inference of the matter density of the Universe, $\Omega_\mathrm{m}$, and the dark energy equation of state, $w$, from the JLA supernovae data set. The prior and exact posterior distribution (obtained from a long MCMC run requiring $\sim 6 \times 10^6$ data model evaluations) are shown in blue and orange, respectively. In the left panel, the approximate rejection-sampling posterior, based on $5,000$ samples accepted out of $\sim 450,000$ simulations, is shown in green. In the right panel, the approximate {\bolfi} posterior, based on $6,000$ simulations only, is shown in red. For all distributions, the $1\sigma$, $2\sigma$ and $3\sigma$ contours are shown. \label{fig:Supernovae}}
\end{center}
\end{figure*}

In this section, we present the first application of {\bolfi} to a cosmological inference problem. Specifically, we perform an analysis of the Joint Lightcurve Analysis (JLA) data set, consisting of the B-band peak apparent magnitudes $m_\mathrm{B}$ of $740$ type Ia supernovae (SN Ia) with redshift $z$ between $0.01$ and $1.3$ \citep{Betoule2014}: $\textbf{d}_\mathrm{O} \equiv \left( m_{\mathrm{B},\mathrm{O}}^k \right)$ for $k \in \llbracket 1,740 \rrbracket$. The details of the data model and inference assumptions are given in appendix \ref{apx:Supernova cosmology}. For the purpose of validating {\bolfi}, we assume a Gaussian synthetic likelihood (see section \ref{sapx:Discrepancy}), allowing us to demonstrate the fidelity of the {\bolfi} posterior against the exact likelihood-based solution obtained via Markov Chain Monte Carlo (MCMC). This analysis can also be compared to the proof of concept for another likelihood-free method, {\delfi} \citep[Density Estimation for Likelihood-Free Inference,][]{Papamakarios2016,Alsing2018}, as the assumptions are very similar.

As described in appendix \ref{apx:Supernova cosmology}, the full problem is six dimensional; however, in this work, we focus on the inference of the two physically relevant quantities, namely $\Omega_\mathrm{m}$ (the matter density of the Universe) and $w$ (the equation of state of dark energy, assumed constant), and marginalise over the other four (nuisance) parameters ($\alpha$, $\beta$, $M_\mathrm{B}$, $\deltaM$). We assume a Gaussian prior,
\begin{equation}
\begin{pmatrix}
\Omega_\mathrm{m} \\
w
\end{pmatrix} \sim
\mathpzc{G}\left[
\begin{pmatrix}
0.3 \\
-0.75
\end{pmatrix},
\begin{pmatrix}
0.4^2 & -0.24 \\
-0.24 & 0.75^2
\end{pmatrix}
\right],
\label{eq:SNe_prior_Omegam_w}
\end{equation}
which is roughly aligned with the direction of the well-known $\Omega_\mathrm{m}-w$ degeneracy. We generated $10^6$ samples (out of $\sim 6\times 10^6$ data model evaluations) of the posterior for the exact six-dimensional Bayesian problem via MCMC \citep[performed using the \textsc{emcee} code,][]{Foreman-Mackey2013}, ensuring sufficient convergence to characterise the $3\sigma$ contours of the distribution.\footnote{The final Gelman-Rubin statistic \citep{Gelman1992} was $R -1 \leq 5 \times 10^{-4}$ for each of the six parameters.} The prior and the exact posterior are shown in blue and orange, respectively, in figure \ref{fig:Supernovae}.

For likelihood-free inference, the simulator takes as input $\Omega_\mathrm{m}$ and $w$ and simulates $N$ realisations of the magnitudes $m_\mathrm{B}$ of the 740 supernovae at their redshifts. Consistently with the Gaussian likelihood used in the MCMC analysis, we assume a Gaussian synthetic likelihood with a fixed covariance matrix $\textbf{C}$. The observed data $\textbf{d}_\mathrm{O}$ and the covariance matrix $\textbf{C}$ are shown in figure \ref{fig:JLA_Hubble_correlation}. 

The approximate posterior obtained from likelihood-free rejection sampling is shown in green in figure \ref{fig:Supernovae}. It was obtained from $5,000$ accepted samples using a (conservative) threshold of $\varepsilon = 650$ on $\Delta_{(\Omega_\mathrm{m},w)}$, chosen so that the acceptance ratio was not below $0.01$. The entire run required $\sim 450,000$ simulations in total. The approximate posterior obtained via {\bolfi} is shown in red in figure \ref{fig:Supernovae}. {\bolfi} was initialised with a Sobol sequence of $20$ samples, and $100$ acquisitions were performed according to the ExpIntVar criterion, without acquisition noise. The {\bolfi} posterior is a much finer approximation to the true posterior than the one obtained from likelihood-free rejection sampling. It is remarkable that only $100$ acquisitions are enough to learn the non-trivial banana shape of the posterior. Only the $3\sigma$ contour \citep[which is usually not shown in cosmology papers, e.g.][]{Betoule2014} notably deviates from the MCMC posterior. This is due to the fact that we used one realisation of the stochastic process defining $\Delta_{(\Omega_\mathrm{m},w)}$ and only $N=50$ realisations per $(\Omega_\mathrm{m},w)$; the marginalisation over the four nuisance parameters is therefore partial, yielding slightly smaller credible contours. However, a better approximation could be obtained straightfowardly, if desired, by investing more computational resources (increasing $N$), without requiring more acquisitions. 

As we used $N=50$, the total cost for {\bolfi} is $6,000$ simulations. This is a reduction by $\sim 2$ orders of magnitude with respect to likelihood-free rejection sampling ($\sim 450,000$ simulations) and $3$ orders of magnitude with respect to MCMC sampling of the exact posterior ($6 \times 10^6$ simulations). It is also interesting to note that our {\bolfi} analysis required a factor of $\sim 3$ fewer simulations than the recently introduced \delfi\ procedure \citep{Alsing2018}, which used $20,000$ simulations drawn from the prior for the analysis of the JLA.\footnote{A notable difference is that {\delfi} allowed the authors to perform the joint inference of the six parameters of the problem, whereas we only get the distribution of $\Omega_\mathrm{m}$ and $w$. However, since these are the only two physically interesting parameters, inference of the nuisance parameters is not deemed crucial for this example.}

\section{Discussion}
\label{sec:Discussion}

\subsection{Benefits and limitations of the proposed approach for cosmological inferences}
\label{ssec:Benefits and limitations of the proposed approach for cosmological inferences}

As noted in the introduction, likelihood-free rejection sampling, when at all viable, is extremely costly in terms of the number of required simulations. In contrast, the {\bolfi} approach relies on a GP probabilistic model for the discrepancy, and therefore allows the incorporation of a smoothness assumption about the approximate likelihood $L(\boldsymbol{\uptheta})$. The smoothness assumption allows simulations in the training set to ``share'' information about their value of $\Delta_{\boldsymbol{\uptheta}}$ in the neighbourhood of $\boldsymbol{\uptheta}$, which suggests that fewer simulations are needed to reach a certain level of accuracy. Indeed, the number of simulations required is typically reduced by $2$ to $3$ orders of magnitude, for a better final approximation of the posterior, as demonstrated by our tests in section \ref{sec:Applications} and in the statistical literature \citep[see][]{GutmannCorander2016}. 

A second benefit of {\bolfi} is that it actively acquires training data through Bayesian optimisation. The trade-off between computational cost and statistical performance is still present, but in a modified form: the trade-off parameter is the size of the training set used in the regression. Within the training set, the user is free to choose which areas of the parameter space should be prioritised, so as to approximate the regression function more accurately there. In contrast, in ABC strategies that rely on drawing from a fixed proposal distribution (often the prior), or variants such as \textsc{pmc}-\textsc{abc}, a fixed computational cost needs to be paid per value of $\boldsymbol{\uptheta}$ regardless of the value of $\Delta_{\boldsymbol{\uptheta}}$. 

Finally, by focusing on parametric approximations to the exact likelihood, the approach proposed in this work is totally ``$\varepsilon$-free'', meaning that no threshold (which is often regarded as an unappealing \textit{ad hock} element) is required. As likelihood-based techniques, the parametric version of {\bolfi} has the drawback that assuming a wrong form for the synthetic likelihood or miscalculating values of its parameters (such as the covariance matrix) can potentially bias the approximate posterior and/or lead to an underestimation of credible regions. Nevertheless, massive data compression procedures can make the assumptions going into the choice of a Gaussian synthetic likelihood (almost) true by construction (see section \ref{sssec:Data compression}).

Of course, regressing the discrepancy and optimising the acquisition function are not free of computational cost. However, the run-time for realistic cosmological simulation models can be hours or days. In comparison, the computational overhead introduced by {\bolfi} is negligible.

Likelihood-free inference should also be compared to existing likelihood-based techniques for cosmology such as Gibbs sampling or Hamiltonian Monte Carlo (e.g. \citealp{Wandelt2004,Eriksen2004} for the cosmic microwave background; \citealp{Jasche2010b,Jasche2015,Jasche2015BORGSDSS} for galaxy clustering; \citealp{Alsing2016} for weak lensing). The principal difference between these techniques and {\bolfi} lies in its likelihood-free nature. Likelihood-free inference has particular appeal for cosmological data analysis, since encoding complex physical phenomena and realistic observational effects into forward simulations is much easier than designing an approximate likelihood which incorporates these effects and solving the inverse problem. While the numerical complexity of likelihood-based techniques typically requires to approximate complex data models in order to access required products (conditionals or gradients of the pdfs) and to allow for sufficiently fast execution speeds, {\bolfi} performs inference from full-scale black-box data models. In the future, such an approach is expected to allow previously infeasible analyses, relying on a much more precise modelling of cosmological data, including in particular the complicated systematics they experience. However, while the physics and instruments will be more accurately modelled, the statistical approximation introduced with respect to likelihood-based techniques should be kept in mind.

Other key aspects of {\bolfi} for cosmological data analysis are the arbitrary choice of the statistical summaries and the easy joint treatment of different data sets. Indeed, as the data compression from $\textbf{d}$ to $\boldsymbol{\Phi}$ is included in the simulator (see section \ref{ssec:Approximate Bayesian computation}), summary statistics do not need to be quantities that can be physically modelled (such as the power spectrum) and can be chosen robustly to model misspecification. For example, for the microwave sky, the summaries could be the cross-spectra between different frequency maps; and for imaging surveys, the cross-correlation between different bands. Furthermore, joint analyses of correlated data sets, which is usually challenging in likelihood-based approaches (as they require a good model for the joint likelihood) can be performed straightforwardly in a likelihood-free approach. 

Importantly, as a general inference technique, {\bolfi} can be embedded into larger probabilistic schemes such as Gibbs or Hamiltonian-within-Gibbs samplers. Indeed, as posterior predictive distributions for conditionals and gradients of GPs are analytically tractable, it is easy to obtain samples of the {\bolfi} approximate posterior for use in larger models. {\bolfi} can therefore allow parts of a larger Bayesian hierarchical model to be treated as black boxes, without compromising the tractability of the entire model. 

\subsection{Possible extensions}
\label{ssec:Possible extensions}

\subsubsection{High-dimensional inference}
\label{sssec:High-dimensional inference}

In this proof-of-concept paper, we focused on two-dimensional problems. Likelihood-free inference is in general very difficult when the dimensionality of the parameter space is large, due to the curse of dimensionality, which makes the volume exponentially larger with $\mathrm{dim}~\boldsymbol{\uptheta}$. In {\bolfi}, this difficulty manifests itself in the form of a hard regression problem which needs to be solved. The areas in the parameter space where the discrepancy is small tend to be narrow in high dimension, therefore discovering these areas becomes more challenging as the dimension increases. The optimisation of GP kernel parameters, which control the shapes of allowed features, also becomes more difficult. Furthermore, finding the global optimum of the acquisition function becomes more demanding (especially with the ones designed for ABC such as ExpIntVar, which have a high degree of structure -- see figure \ref{fig:Supernovae_acquisition}, bottom right panel).

Nevertheless, \citet{Jaervenpaeae2017} showed on a toy simulation model (a Gaussian) that up to ten-dimensional inference is possible with {\bolfi}. As usual cosmological models do not include more than ten free physical parameters, we do not expect this limitation to be a hindrance. Any additional nuisance parameter or latent variable used internally by the simulator (such as $\alpha$, $\beta$, $M_\mathrm{B}$, $\deltaM$ in supernova cosmology, see section \ref{ssec:Supernova cosmology}) can be automatically marginalised over, by using $N$ realisations per $\boldsymbol{\uptheta}$. Recent advances in high-dimensional implementation of the synthetic likelihood \citep{Ong2017} and high-dimensional Bayesian optimisation \citep[e.g.][]{Wang2013:BOH:2540128.2540383,Kandasamy2015} could also be exploited. In future work, we will address the problem of high-dimensional likelihood-free inference in a cosmological context.

\subsubsection{Scalability with the number of acquisitions and probabilistic model for the discrepancy}
\label{sssec:Scalability with the number of acquisitions and probabilistic model for the discrepancy}

In addition to the fundamental issues with high-dimensional likelihood-free inference described in the previous section, practical difficulties can be met.

Gaussian process regression requires the inversion of a matrix $\uuline{\textbf{K}}\vspace{-4pt}$ of size $t \times t$, where $t$ is the size of the training set. The complexity is $\mathcal{O}(t^3)$, which limits the size of the training set to a few thousand. Improving GPs with respect to this inversion is still subject to research \citep[see][chapter 8]{RasmussenWilliams2006}. For example, ``sparse'' Gaussian process regression reduces the complexity by introducing auxiliary ``inducing variables''. Techniques inspired by the solution to the Wiener filtering problem in cosmology, such as preconditioned conjugate gradient or messenger field algorithms could also be used \citep{Elsner2013,KodiRamanah2017,Papez2018}. Another strategy would be to divide the regression problem spatially into several patches with a lower number of training points \citep{Park2017}. Such approaches are possible extensions of the presented method.

In the GP probabilistic model employed to model the discrepancy, the variance depends only on the training locations, not on the obtained values (see equation \eqref{eq:GP_variance}). Furthermore, a stationary kernel is assumed. However, depending on the simulator, the discrepancy can show heteroscedasticity (i.e. its variance can depend on $\boldsymbol{\uptheta}$ -- see e.g. figure \ref{fig:Gaussian_mean_illustration}, bottom panel). Such cases could be handled by non-stationary GP kernels or different probabilistic models for the discrepancy, allowing a heteroscedastic regression.

\subsubsection{Acquisition rules}
\label{sssec:Acquisition rules}

As shown in our examples, attention should be given to the selection of an efficient acquisition rule. Although standard Bayesian optimisation strategies such as the EI are reasonably effective, they are usually too greedy, focusing nearly all the sampling effort near the estimated minimum of the discrepancy and gathering too little information about other regions in the domain (see figure \ref{fig:Supernovae_acquisition}, bottom left panel). This implies that, unless the acquisition noise is high, the tails of the posterior will not be as well approximated as the modal areas. In contrast, the ExpIntVar acquisition rule, derived in this work for the parametric approach, addresses the inefficient use of resources in likelihood-free rejection sampling by directly targeting the regions of the parameter space where improvement in the estimation accuracy of the approximate posterior is needed most. In our experiments, ExpIntVar seems to correct -- at least partially -- for the well-known effect in Bayesian optimisation of overexploration of the domain boundaries, which becomes more problematic in high dimension.

Acquisition strategies examined so far in the literature \citep[see][for a comparative study]{Jaervenpaeae2017} have focused on single acquisitions and are all ``myopic'', in the sense that they reason only about the expected utility of the next acquisition, and the number of simulations left in a limited budget is not taken into account. Improvement of acquisition rules enabling batch acquisitions and non-myopic reasoning are left to future extensions of {\bolfi}.

\subsubsection{Data compression}
\label{sssec:Data compression}

In addition to the problem of the curse of dimensionality in parameter space, discussed in section \ref{sssec:High-dimensional inference}, likelihood-free inference usually suffers from difficulties in the measuring the (mis)match between simulations and observations if the data space also has high dimension. As discussed in section \ref{ssec:Approximate Bayesian computation}, simulator-based models include a data compression step. The comparison in data space can be made more easily if $\mathrm{dim}~\boldsymbol{\Phi}$ is reduced. In future work, we will therefore aim at combining {\bolfi} with massive and (close to) optimal data compression strategies. These include \textsc{moped} \citep{Heavens2000}, the score function \citep{AlsingWandelt2018}, or information-maximising neural networks \citep{Charnock2018}. Using such efficient data compression techniques, the number of simulations required for inference with {\bolfi} will be reduced even more, and the number of parameters treated could be increased.

Parametric approximations to the exact likelihood depend on quantities that have to be estimated using the simulator (typically for the Gaussian synthetic likelihood, the inverse covariance matrix of the summaries). Unlike supernova cosmology where the covariance matrix is easily obtained, in many cases it is prohibitively expensive to run enough simulations to estimate the required quantities, especially when they vary with the model parameters. In this context, massive data compression offers a way forward, reducing enormously the number of required simulations and making the analysis feasible when otherwise it might be essentially impossible \citep{Heavens2017,Gualdi2018}.

An additional advantage of several data compression strategies is that they support the choice of a Gaussian synthetic likelihood. Indeed, the central limit theorem (for \textsc{moped}) or the form of the network's reward function (for information-maximising neural networks) assist in giving the compressed data a near-Gaussian distribution. Furthermore, testing the Gaussian assumption for the synthetic likelihood will be far easier in a smaller number of dimensions than in the original high-dimensional data space.

\subsection{Parallelisation and computational efficiency}
\label{ssec:Parallelisation and computational efficiency}

While MCMC sampling has to be done sequentially, {\bolfi} lends itself to more parallelisation. In an efficient strategy, a master process performs the regression and decides on acquisition locations, then dispatches simulations to be run by different workers. In this way, many simulations can be run simultaneously in parallel, or even on different machines. This allows fast application of the method and makes it particularly suitable for grid computing. Extensions of the probabilistic model and of the acquisition rules, discussed in section \ref{sssec:Scalability with the number of acquisitions and probabilistic model for the discrepancy} and \ref{sssec:Acquisition rules}, would open the possibility of doing asynchronous acquisitions. Different workers would then work completely independently and decide on their acquisitions locally, while just sharing a pool of simulations to update their beliefs given all the evidence available.

While the construction of the training set depends on the observed data $\boldsymbol{\Phi}_\mathrm{O}$ (through the acquisition function), simulations can nevertheless be reused as long as summaries $\boldsymbol{\Phi}_{\boldsymbol{\uptheta}}$ are saved. This means that if one acquires new data $\boldsymbol{\Phi}_\mathrm{O}'$, the existing $\boldsymbol{\Phi}_{\boldsymbol{\uptheta}}$ (or a subset of them) can be used to compute the new discrepancy $\Delta_{\boldsymbol{\uptheta}}(\boldsymbol{\Phi}_{\boldsymbol{\uptheta}}, \boldsymbol{\Phi}_\mathrm{O}')$. Building an initial training set in this fashion can massively speed up the inference of $\p(\boldsymbol{\uptheta}|\boldsymbol{\Phi})_{\boldsymbol{\Phi}=\boldsymbol{\Phi}_\mathrm{O}'}$, whereas likelihood-based techniques would require a new MCMC.

\subsection{Comparison to previous work}
\label{ssec:Comparison to previous work}

As discussed in the introduction, likelihood-free rejection sampling is not a viable strategy for various problems that {\bolfi} can tackle. In recent work, an other algorithm for scalable likelihood-free inference in cosmology \citep[{\delfi},][]{Papamakarios2016,Alsing2018} was introduced. The approach relies on estimating the joint probability $\p(\boldsymbol{\uptheta},\boldsymbol{\Phi})$ via density estimation. This idea also relates to the work of \citet{Hahn2018}, who fit the sampling distribution of summaries $\p(\boldsymbol{\Phi}|\boldsymbol{\uptheta})$ using Gaussian mixture density estimation or independent component analysis, before using it for parameter estimation. This section discusses the principal similarities and differences.

The main difference between {\bolfi} and {\delfi} is the data acquisition. Training data are actively acquired in {\bolfi}, contrary to {\delfi} which, in the simplest scheme, draws from the prior. The reduction in the number of simulations for the inference of cosmological parameters (see section \ref{ssec:Supernova cosmology}) can be interpreted as the effect of the Bayesian optimisation procedure in combination with the ExpIntVar acquisition function. Using a purposefully constructed surrogate surface instead of a fixed proposal distribution, {\bolfi} focuses the simulation effort to reveal as much information as possible about the target posterior. In particular, its ability to reason about the quality of simulations before they are run is an essential element. Acquisition via Bayesian optimisation almost certainly remains more efficient than even the \textsc{pmc} version of {\delfi}, which learns a better proposal distribution but still chooses parameters randomly. In future cosmological applications with simulators that are expensive and/or have a large latent space, an active data acquisition procedure could be crucial in order to provide a good model for the noisy approximate likelihood in the interesting regions of parameter space, and to reduce the computational cost. This comes at the expense of a reduction of the parallelisation potential: with a fixed proposal distribution (like in {\delfi} and unlike in {\bolfi}), the entire set of simulations can be run at the same time.

The second comment is related to the dimensionality of problems which can be addressed. Like {\delfi}, {\bolfi} relies on a probabilistic model to make ABC more efficient. However, the quantities employed differ, since in {\delfi} the relation between the parameters $\boldsymbol{\uptheta}$ and the summary statistics $\boldsymbol{\Phi}$ is modelled (via density estimation), while {\bolfi} focuses on the relation between the parameters $\boldsymbol{\uptheta}$ and the discrepancy $\Delta_{\boldsymbol{\uptheta}}$ (via regression). Summary statistics are multi-dimensional while the discrepancy is a univariate scalar quantity. Thus, {\delfi} requires to solve a density estimation problem in $\mathrm{dim}~\boldsymbol{\uptheta} + \mathrm{dim}~\boldsymbol{\Phi}$ (which equals $2 \times \mathrm{dim}~\boldsymbol{\uptheta}$ if the compression from \citealp{AlsingWandelt2018} is used), while {\bolfi} requires to solve a regression problem in $\mathrm{dim}~\boldsymbol{\uptheta}$. Both tasks are expected to become more difficult as $\mathrm{dim}~\boldsymbol{\uptheta}$ increases (a symptom of the curse of dimensionality, see section \ref{sssec:High-dimensional inference}), but the upper limits on $\mathrm{dim}~\boldsymbol{\uptheta}$ for practical applications may differ. Further investigations are required to compare the respective maximal dimensions of problems that can be addressed by {\bolfi} and {\delfi}.

Finally, as argued by \citet{Alsing2018}, {\delfi} readily provides an estimate of the approximate evidence. In contrast, as in likelihood-based techniques, integration over parameter space is required with {\bolfi} to get
\begin{equation}
Z_{\boldsymbol{\Phi}} = \left(\int \p(\boldsymbol{\Phi}|\boldsymbol{\uptheta}) \, \mathrm{d}\boldsymbol{\uptheta} \right)_{\boldsymbol{\Phi}=\boldsymbol{\Phi}_\mathrm{O}}.
\end{equation}
However, due to the GP model, the integral can be more easily computed, using the same strategies as for the integral appearing in ExpIntVar (see section \ref{sssec:Expected integrated variance}): only the GP predicted values are required at discrete locations on a grid (in low dimension) or at the positions of importance samples. A potential caveat is that {\delfi} has only been demonstrated to work in combination with the score function \citep{AlsingWandelt2018}, which is necessary to reduce the dimensionality of $\boldsymbol{\Phi}$ before estimating the density.\footnote{In contrast, section \ref{ssec:Supernova cosmology} showed, for the same supernovae problem, that {\bolfi} can still operate if the comparison is done in the full $740$-dimensional data space.} The score function produces summaries that are only sufficient up to linear order in the log-likelihood. However, in ABC, care is required to perform model selection if the summary statistics are insufficient. Indeed, \citet[][equation 1]{Robert2011} show that, in such a case, the approximate Bayes factor can be arbitrarily biased and that the approximation error is unrelated to the computational effort invested in running the ABC algorithm. Moreover, sufficiency for models $\mathcal{M}_1$ and $\mathcal{M}_2$ alone, or even for both of them -- even if approximately realised via Alsing \& Wandelt's procedure -- does not guarantee sufficiency to compare the two different models $\mathcal{M}_1$ and $\mathcal{M}_2$ \citep{Didelot2011}. As the assumptions behind {\bolfi} do not necessarily necessitate to reduce $\mathrm{dim}~\boldsymbol{\Phi}$ ($\Delta_{\boldsymbol{\uptheta}}$ is always a univariate scalar quantity, see above), these difficulties could be alleviated with {\bolfi} by carefully designing sufficient summary statistics for model comparison within the black-box simulator, if they exist.

\section{Conclusion}
\label{sec:Conclusion}

Likelihood-free inference methods allow Bayesian inference of the parameters of simulator-based statistical models with no reference to the likelihood function. This is of particular interest for data analysis in cosmology, where complex physical and observational processes can usually be simulated forward but not handled in the inverse problem. 

In this paper, we considered the demanding problem of performing Bayesian inference when simulating data from the model is extremely costly. We have seen that likelihood-free rejection sampling suffers from a vanishingly small acceptance rate when the threshold $\varepsilon$ goes to zero, leading to the need for a prohibitively large number of simulations. This high cost is largely due to the lack of knowledge about the functional relation between the model parameters and the discrepancy. As a response, we have described a new approach to likelihood-free inference, {\bolfi}, that uses regression to infer this relation, and optimisation to actively build the training data set. A crucial ingredient is the acquisition function derived in this work, with which training data are acquired such that the expected uncertainty in the final estimate of the posterior is minimised.

In case studies, we have shown that {\bolfi} is able to precisely recover the true posterior, even far in its tails, with as few as $6,000$ simulations, in contrast to likelihood-free rejection sampling or likelihood-based MCMC techniques which require orders of magnitude more simulations. The reduction in the number of required simulations accelerated the inference massively.

This study opens up a wide range of possible extensions, discussed in section \ref{ssec:Possible extensions}. It also allows for novel analyses of cosmological data from fully non-linear simulator-based models, as required e.g. for the cosmic web \citep[see the discussions in][]{Leclercq2015ST,Leclercq2016CIT,Leclercq2017DMSHEET}. Other applications may include the cosmic microwave background, weak gravitational lensing or intensity mapping experiments. We therefore anticipate that {\bolfi} will be a major ingredient in principled, simulator-based inference for the coming era of massive cosmological data.

\appendix
\counterwithin{figure}{section}

\section{Derivations of the mathematical results}
\label{apx:Derivations of the mathematical results}

\subsection{Expressions for the approximate posterior}
\label{sapx:Expressions for the approximate posterior}

If we knew the target function $f$, the {\bolfi} posterior would be given as
\begin{equation}
\p_{\bolfi}(\boldsymbol{\uptheta}|\boldsymbol{\Phi}_\mathrm{O}) \equiv \p(\boldsymbol{\uptheta}) \exp\left( -\frac{1}{2} f(\boldsymbol{\uptheta}) \right) \propto \p(\boldsymbol{\uptheta}) \exp\left( \tilde{\ell}(\boldsymbol{\uptheta}) \right).
\end{equation}
However, due to the limited computational resources we only have a finite training set $(\boldsymbol{\Theta}, \textbf{f})$, which implies that there is uncertainty in the values of $f(\boldsymbol{\uptheta})$, and therefore that the approximate posterior is itself a stochastic process. To get its expectation under the model, the log-likelihood $\tilde{\ell}(\boldsymbol{\uptheta})$ is replaced by its expectation under the model, i.e. $-\frac{1}{2}\mu(\boldsymbol{\uptheta})$ (up to constants, see equations \eqref{eq:l_J_proposition_1} and \eqref{eq:J_t_equals_mu}), giving equation \eqref{eq:approximate_posterior_expectation}.

Similarly, if the function $f$ was known, the variance of the approximate posterior could be computed by standard propagation of uncertainties,
\begin{eqnarray}
\mathrm{V}\left[ \p_{\bolfi}(\boldsymbol{\uptheta}|\boldsymbol{\Phi}_\mathrm{O}, \textbf{f}, \boldsymbol{\Theta}) \right] & = & \left| \frac{\partial}{\partial f} \p(\boldsymbol{\uptheta}) \exp\left( -\frac{1}{2} f \right) \right|^2 \mathrm{V}\left[ f \right] \nonumber \\
& = & \frac{\p(\boldsymbol{\uptheta})^2}{4} \exp(-f) \mathrm{V}\left[ f \right] .
\end{eqnarray}
The argument of the exponential is $-f(\boldsymbol{\uptheta}) = 2 \tilde{\ell}(\boldsymbol{\uptheta})$; it should be replaced by its expectation under the model, $-\mu(\boldsymbol{\uptheta})$. The variance of $f$ under the model is, by definition, $\mathrm{V}^{(\mathrm{t})}\left[ f \right] = \sigma^2(\boldsymbol{\uptheta})$. The result for $\mathrm{V}^{(\mathrm{t})}\left[ \p_{\bolfi}(\boldsymbol{\uptheta}|\boldsymbol{\Phi}_\mathrm{O}, \textbf{f}, \boldsymbol{\Theta}) \right]$ is therefore given by equation \eqref{eq:approximate_posterior_variance}.

\subsection{The ExpIntVar acquisition function in the parametric approach}
\label{sapx:The ExpIntVar acquisition function in the parametric approach}

We start by deriving the probability distributions for the GP mean and variance after one future observation $(\boldsymbol{\uptheta}_\star, f_\star)$ is added to the training set $(\boldsymbol{\Theta}, \textbf{f})$. We denote them by $\mu_\star$ and $\sigma^2_\star$ respectively. These quantities are random functions of $\boldsymbol{\uptheta}$ since the new value $f_\star$ is unknown. Assuming that the GP mean is $m(\boldsymbol{\uptheta})=0$ for simplicity, and using equation \eqref{eq:GP_mean} with the full training set $\left\lbrace(\boldsymbol{\Theta}, \textbf{f}), (\boldsymbol{\uptheta}_\star, f_\star)\right\rbrace$, we get
\begin{equation}
\mu_\star(\boldsymbol{\uptheta}) = \begin{pmatrix}
\uline{\textbf{K}} \\
\kappa(\boldsymbol{\uptheta},\boldsymbol{\uptheta}_\star)
\end{pmatrix}^\intercal \begin{pmatrix}
\uuline{\textbf{K}} & \uline{\textbf{K}}_\star \\
\uline{\textbf{K}}_\star^\intercal & K_{\star\star}
\end{pmatrix}^{-1} \begin{pmatrix}
\textbf{f} \\
f_\star
\end{pmatrix},
\end{equation}
using the notations of equations \eqref{eq:GP_notation_def_1}--\eqref{eq:GP_notation_def_4} and \eqref{eq:GP_notation_def_5}. By means of a standard formula for block matrix inversion, we get
\begin{eqnarray}
\mu_\star(\boldsymbol{\uptheta}) & = & \uline{\textbf{K}}^\intercal \uuline{\textbf{K}}^{-1}\textbf{f} + \left[ \kappa(\boldsymbol{\uptheta},\boldsymbol{\uptheta}_\star) - \uline{\textbf{K}}^\intercal \uuline{\textbf{K}}^{-1} \uline{\textbf{K}}_\star \right] \times \nonumber\\
& & \left[ K_{\star\star} - \uline{\textbf{K}}_\star^\intercal \uuline{\textbf{K}}^{-1} \uline{\textbf{K}}_\star \right]^{-1} \left[ f_\star - \uline{\textbf{K}}_\star^\intercal \uuline{\textbf{K}}^{-1} \textbf{f} \right] \\
& = & \mu(\boldsymbol{\uptheta}) + \mathrm{cov}(\boldsymbol{\uptheta},\boldsymbol{\uptheta}_\star) \times \left[\sigma^2(\boldsymbol{\uptheta}_\star)\right]^{-1} \left[ f_\star - \mu(\boldsymbol{\uptheta}_\star) \right]. \nonumber\label{eq:GP_new_mean}
\end{eqnarray}
According to the GP model trained with $\left\lbrace(\boldsymbol{\Theta}, \textbf{f})\right\rbrace$, the unknown future observation $f_\star$ is Gaussian-distributed, i.e. $\p(f_\star|\textbf{f},\boldsymbol{\Theta},\boldsymbol{\uptheta}_\star) = \mathpzc{G}(\mu(\boldsymbol{\uptheta}_\star), \sigma^2(\boldsymbol{\uptheta}_\star))$. Thus, $\left[\sigma^2(\boldsymbol{\uptheta}_\star)\right]^{-1} \left[ f_\star - \mu(\boldsymbol{\uptheta}_\star) \right]$ is Gaussian-distributed with mean zero and variance $\left[\sigma^2(\boldsymbol{\uptheta}_\star)\right]^{-1}$, and $\mu_\star(\boldsymbol{\uptheta})$ is Gaussian-distributed with mean $\mu(\boldsymbol{\uptheta})$ and variance $\tau^2(\boldsymbol{\uptheta},\boldsymbol{\uptheta}_\star)$, 
\begin{equation}
\p(\mu_\star(\boldsymbol{\uptheta})|\textbf{f}, \boldsymbol{\Theta}, \boldsymbol{\uptheta}_\star) = \mathpzc{G}(\mu(\boldsymbol{\uptheta}), \tau^2(\boldsymbol{\uptheta},\boldsymbol{\uptheta}_\star)),
\end{equation}
using the notation introduced in equation \eqref{eq:def_tau}.

Similar calculations for the variance show that
\begin{equation}
\sigma^2_\star(\boldsymbol{\uptheta}) = \sigma^2(\boldsymbol{\uptheta}) - \tau^2(\boldsymbol{\uptheta},\boldsymbol{\uptheta}_\star),
\label{eq:GP_new_variance}
\end{equation}
and therefore
\begin{equation}
\p(\sigma^2_\star(\boldsymbol{\uptheta})|\textbf{f}, \boldsymbol{\Theta}, \boldsymbol{\uptheta}_\star) = \updelta_\mathrm{D}\left( \sigma^2_\star(\boldsymbol{\uptheta}) - \sigma^2(\boldsymbol{\uptheta}) + \tau^2(\boldsymbol{\uptheta},\boldsymbol{\uptheta}_\star) \right).
\end{equation}
This formula means that the reduction in the GP variance is deterministic and depends only on the new location $\boldsymbol{\uptheta}_\star$, independently of the future observation $f_\star$.

We now derive the expression for the expected integrated variance in the parametric approach.
\begin{eqnarray}
&&\!\!\!\!\!\!\!\!\mathrm{EIV}(\boldsymbol{\uptheta}_\star) \equiv \mathrm{E}^{(\mathrm{t})} \left[ \mathpzc{L}\left[ \p_{\bolfi}(\boldsymbol{\uptheta}|\boldsymbol{\Phi}_\mathrm{O}, \textbf{f}, \boldsymbol{\Theta}, f_\star, \boldsymbol{\uptheta}_\star) \right] \right] \nonumber\\
& = & \int \mathpzc{L}\left[ \p_{\bolfi}(\boldsymbol{\uptheta}|\boldsymbol{\Phi}_\mathrm{O}, \textbf{f}, \boldsymbol{\Theta}, f_\star, \boldsymbol{\uptheta}_\star) \right] \p(f_\star|\textbf{f}, \boldsymbol{\Theta}, \boldsymbol{\uptheta}_\star) \,  \mathrm{d}f_\star \nonumber\\
& = & \iint \mathrm{V}\left[ \p_{\bolfi}(\boldsymbol{\uptheta}|\boldsymbol{\Phi}_\mathrm{O},\textbf{f},\boldsymbol{\Theta}) \right] \mathrm{d}\boldsymbol{\uptheta} \, \p(f_\star|\textbf{f}, \boldsymbol{\Theta}, \boldsymbol{\uptheta}_\star) \, \mathrm{d}f_\star \nonumber\\
& = & \int \p(\boldsymbol{\uptheta})^2 w^2(\boldsymbol{\uptheta},\boldsymbol{\uptheta}_\star) \, \mathrm{d}\boldsymbol{\uptheta}, 
\end{eqnarray}
where in the last line we have interchanged the order of integration, used equation \eqref{eq:approximate_posterior_variance}, and introduced 
\begin{equation}
\begin{split}
w^2(\boldsymbol{\uptheta},\boldsymbol{\uptheta}_\star) & \equiv \int \frac{1}{4} \exp\left[-\mu_\star(\boldsymbol{\uptheta}) \right] \sigma^2_\star(\boldsymbol{\uptheta}) \, \p(f_\star|\textbf{f}, \boldsymbol{\Theta}, \boldsymbol{\uptheta}_\star) \, \mathrm{d}f_\star \\
& = \mathrm{E}^{(\mathrm{t})} \left[ \frac{1}{4} \exp\left[-\mu_\star(\boldsymbol{\uptheta}) \right] \sigma^2_\star(\boldsymbol{\uptheta}) \right],
\end{split}
\end{equation}
that is to say the expectation of $\frac{1}{4} \exp\left[-\mu_\star(\boldsymbol{\uptheta}) \right] \sigma^2_\star(\boldsymbol{\uptheta})$ under the GP model trained with $\left\lbrace(\boldsymbol{\Theta}, \textbf{f}), (\boldsymbol{\uptheta}_\star, f_\star)\right\rbrace$. This expectation can be treated using equations \eqref{eq:GP_new_mean} and \eqref{eq:GP_new_variance}, assuming that mean and variance are independent: $\sigma^2_\star(\boldsymbol{\uptheta})$ becomes deterministically $\sigma^2(\boldsymbol{\uptheta}) - \tau^2(\boldsymbol{\uptheta},\boldsymbol{\uptheta}_\star)$ under the model. As in section \ref{sapx:Expressions for the approximate posterior}, the argument of the exponential, $\mu_\star(\boldsymbol{\uptheta})$, is replaced by its mean $\mu(\boldsymbol{\uptheta})$. The final result is
\begin{equation}
w^2(\boldsymbol{\uptheta},\boldsymbol{\uptheta}_\star) = \frac{1}{4} \exp\left[ -\mu(\boldsymbol{\uptheta}) \right] \left[ \sigma^2(\boldsymbol{\uptheta})-\tau^2(\boldsymbol{\uptheta},\boldsymbol{\uptheta}_\star) \right].
\end{equation}

\subsection{Gradient of the ExpIntVar acquisition function in the parametric approach}
\label{sapx:Gradient of the ExpIntVar acquisition function in the parametric approach}

In this section we derive the gradient of the expected integrated variance in the parametric approach, which can be used to find its minimum in parameter space. Inverting the differentiation and the integration, we have
\begin{eqnarray}
\frac{\mathrm{d} \, \mathrm{EIV}(\boldsymbol{\uptheta}_\star)}{\mathrm{d}\boldsymbol{\uptheta}_\star} & = & \frac{\mathrm{d}}{\mathrm{d}\boldsymbol{\uptheta}_\star} \int \p(\boldsymbol{\uptheta})^2 w^2(\boldsymbol{\uptheta},\boldsymbol{\uptheta}_\star) \, \mathrm{d}\boldsymbol{\uptheta} \nonumber\\
& = & \int \p(\boldsymbol{\uptheta})^2 \, \frac{\partial w^2(\boldsymbol{\uptheta},\boldsymbol{\uptheta}_\star)}{\partial \boldsymbol{\uptheta}_\star} \, \mathrm{d}\boldsymbol{\uptheta},
\label{eq:EIV_gradient}
\end{eqnarray}
where
\begin{eqnarray}
\frac{\partial w^2(\boldsymbol{\uptheta},\boldsymbol{\uptheta}_\star)}{\partial \boldsymbol{\uptheta}_\star} & = & \frac{\partial}{\partial \boldsymbol{\uptheta}_\star} \left\lbrace \frac{1}{4} \exp\left[ -\mu(\boldsymbol{\uptheta}) \right] \left[ \sigma^2(\boldsymbol{\uptheta})-\tau^2(\boldsymbol{\uptheta},\boldsymbol{\uptheta}_\star) \right] \right\rbrace \nonumber\\
& = & -\frac{1}{4} \exp\left[ -\mu(\boldsymbol{\uptheta}) \right] \frac{\partial \tau^2(\boldsymbol{\uptheta},\boldsymbol{\uptheta}_\star)}{\partial \boldsymbol{\uptheta}_\star},
\end{eqnarray}
with
\begin{eqnarray}
\frac{\partial \tau^2(\boldsymbol{\uptheta},\boldsymbol{\uptheta}_\star)}{\partial \boldsymbol{\uptheta}_\star} & = & 2 \frac{\mathrm{cov}(\boldsymbol{\uptheta},\boldsymbol{\uptheta}_\star)}{\sigma^2(\boldsymbol{\uptheta}_\star)} \frac{\partial \, \mathrm{cov}(\boldsymbol{\uptheta},\boldsymbol{\uptheta}_\star)}{\partial \boldsymbol{\uptheta}_\star} \nonumber\\
& & - \frac{\mathrm{cov}(\boldsymbol{\uptheta},\boldsymbol{\uptheta}_\star)}{\sigma^4(\boldsymbol{\uptheta}_\star)} \frac{\partial \sigma^2(\boldsymbol{\uptheta})}{\partial \boldsymbol{\uptheta}_\star} ,\\
\frac{\partial \, \mathrm{cov}(\boldsymbol{\uptheta},\boldsymbol{\uptheta}_\star)}{\partial \boldsymbol{\uptheta}_\star} & = & \frac{\partial \kappa(\boldsymbol{\uptheta},\boldsymbol{\uptheta}_\star)}{\partial \boldsymbol{\uptheta}_\star} - \uline{\textbf{K}}^\intercal \uuline{\textbf{K}}^{-1} \frac{\partial \uline{\textbf{K}}_\star}{\partial \boldsymbol{\uptheta}_\star}.
\end{eqnarray}
The integral in equation \eqref{eq:EIV_gradient} can be evaluated similarly as discussed in section \ref{sssec:Expected integrated variance}.

\section{Summarising Gaussian signals}
\label{apx:Summarising Gaussian signals}

This appendix gives the details of the problem of summarising Gaussian signals discussed in section \ref{ssec:Summarising Gaussian signals}. 

\subsection{Forward modelling}
\label{sapx:Forward modelling}

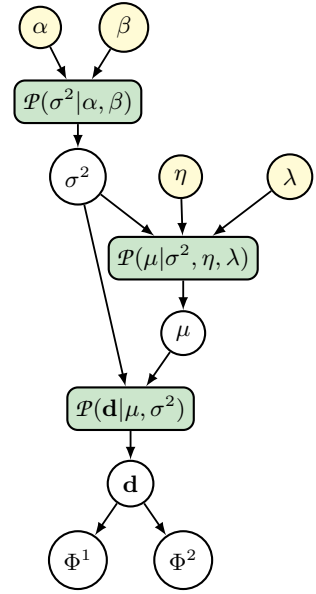
\begin{figure}
\begin{center}
\begin{tikzpicture}
	\pgfdeclarelayer{background}
	\pgfdeclarelayer{foreground}
	\pgfsetlayers{background,main,foreground}

	\tikzstyle{probability}=[draw, thick, text centered, rounded corners, minimum height=1em, minimum width=1em, fill=green!20]
	\tikzstyle{variabl}=[draw, thick, text centered, circle, minimum height=1em, minimum width=1em]
	\tikzstyle{hyperparam}=[draw, thick, text centered, circle, minimum height=1em, minimum width=1em, fill=yellow!20]

	\def\blockdist{0.7}
	\def\shiftdist{0.7}

    \node (alpha) [hyperparam]
    {$\alpha$};
    \path (alpha.west)+(2*\blockdist,0) node (beta) [hyperparam]
    {$\beta$};
    \path (alpha.south)+(0.5,-\blockdist) node (sigmaproba) [probability]
    {$\p(\sigma^2|\alpha, \beta)$};
    \path (sigmaproba.south)+(0,-\blockdist) node (sigma) [variabl]
    {$\sigma^2$};
	\path (sigma.west)+(2.5*\blockdist,0) node (eta) [hyperparam]
	{$\eta$};    
	\path (eta.west)+(2.5*\blockdist,0) node (lambda) [hyperparam]
	{$\lambda$};    
    \path (sigma.south)+(2*\shiftdist,-\blockdist) node (muproba) [probability]
    {$\p(\mu|\sigma^2, \eta, \lambda)$};
    \path (muproba.south)+(0,-\blockdist) node (mu) [variabl]
    {$\mu$};
    \path (mu.south)+(-\shiftdist,-\blockdist) node (dproba) [probability]
    {$\p(\textbf{d}|\mu, \sigma^2)$};
    \path (dproba.south)+(0,-\blockdist) node (d) [variabl]
    {$\textbf{d}$};
    \path (d.south)+(-\shiftdist,-\blockdist) node (phi1) [variabl]
    {$\Phi^1$};
    \path (d.south)+(\shiftdist,-\blockdist) node (phi2) [variabl]
    {$\Phi^2$};

	\draw[line width=0.7pt, arrows={-latex}] (alpha) -- (sigmaproba);
	\draw[line width=0.7pt, arrows={-latex}] (beta) -- (sigmaproba);
	\draw[line width=0.7pt, arrows={-latex}] (sigmaproba) -- (sigma);
	\draw[line width=0.7pt, arrows={-latex}] (sigma) -- (muproba);
	\draw[line width=0.7pt, arrows={-latex}] (eta) -- (muproba);
	\draw[line width=0.7pt, arrows={-latex}] (lambda) -- (muproba);
	\draw[line width=0.7pt, arrows={-latex}] (muproba) -- (mu);
	\draw[line width=0.7pt, arrows={-latex}] (sigma) -- (dproba);
	\draw[line width=0.7pt, arrows={-latex}] (mu) -- (dproba);
	\draw[line width=0.7pt, arrows={-latex}] (dproba) -- (d);
	\draw[line width=0.7pt, arrows={-latex}] (d) -- (phi1);
	\draw[line width=0.7pt, arrows={-latex}] (d) -- (phi2);

\end{tikzpicture}
\end{center}
\caption{Hierarchical forward model for the problem of summarising simulated Gaussian signals. The upper part corresponds to the generation of random variables from the two-dimensional Gaussian-inverse-Gamma prior parametrised by $(\alpha, \beta, \eta, \lambda)$: first $\sigma^2$ is drawn from $\p(\sigma^2|\alpha, \beta)$ (an inverse-Gamma distribution with shape parameter $\alpha$ and scale parameter $\beta$), then $\mu$ is drawn from $\p(\mu|\sigma^2, \eta, \lambda)$ (a Gaussian distribution with mean $\eta$ and variance $\sigma^2/\lambda$). A Gaussian likelihood $\p(\textbf{d}|\mu, \sigma^2)$ with mean $\mu$ and variance $\sigma^2$ gives the data $\textbf{d}$. Finally, the simulator produces two summary statistics: the estimated mean and variance, $\Phi^1$ and $\Phi^2$ respectively.\label{fig:BHM_Gaussian_model}}
\end{figure}

The problem considered is the joint inference of the mean $\mu$ and of the variance $\sigma^2$ of a Gaussian $\mathpzc{G}$, from which we have $n$ samples that constitute the observed data $\textbf{d}_\mathrm{O}$. The true likelihood for this problem is therefore
\begin{equation}
\mathcal{L}(\mu, \sigma^2) \equiv \p(\textbf{d}|\mu,\sigma^2)_{|\textbf{d}=\textbf{d}_\mathrm{O}} = \mathpzc{G}(\textbf{d}|\mu,\sigma^2)_{|\textbf{d}=\textbf{d}_\mathrm{O}}.
\label{eq:Gaussian_true_likelihood}
\end{equation}

The Gaussian-inverse-Gamma is the natural prior for this problem, as it is conjugate for the Gaussian distribution with unknown mean and variance. It is a two-dimensional distribution characterised by four hyperparameters $(\alpha, \beta, \eta, \lambda)$. Samples of this prior can be straightforwardly generated by first sampling $\sigma$ from the inverse-Gamma distribution $\varGamma^{-1}$ with shape parameter $\alpha$ and scale parameter $\beta$, then by drawing $\mu$ from the Gaussian distribution $\mathpzc{G}$ with mean $\eta$ and variance $\sigma^2/\lambda$. 

A noise-free simulator can be designed for this inference problem by taking the operations successively
\begin{eqnarray}
\sigma^2 & \curvearrowleft & \p(\sigma^2|\alpha,\beta) = \varGamma^{-1}(\sigma^2|\alpha,\beta), \\
\mu & \curvearrowleft & \p(\mu|\sigma^2,\eta,\lambda) = \mathpzc{G}(\mu|\eta,\sigma^2/\lambda), \\
\textbf{d} & \curvearrowleft & \p(\textbf{d}|\mu,\sigma^2) = \mathpzc{G}(\textbf{d}|\mu,\sigma^2) .
\end{eqnarray}
After the full data $\textbf{d}$ are generated, they can be compressed to summary statistics. A simple choice is the empirical estimator for the mean and (unbiased) variance, defined by
\begin{eqnarray}
\Phi^1(\textbf{d}) & = & \frac{1}{n} \sum_{k=1}^n d_k, \label{eq:Gaussian_def_Phi1} \\
\Phi^2(\textbf{d}) & = & \frac{1}{n-1} \sum_{k=1}^n \left( d_k - \Phi^1(\textbf{d}) \right)^2 . \label{eq:Gaussian_def_Phi2} 
\end{eqnarray}
$\boldsymbol{\Phi}=(\Phi^1,\Phi^2)$ is a sufficient summary statistic for the inference of $(\mu, \sigma^2)$. For this model, no information is lost in the reduction from $\textbf{d}$ to $\boldsymbol{\Phi}$, which ensures $L(\boldsymbol{\theta}) \propto \mathcal{L}(\boldsymbol{\theta})$. Furthermore, the distribution of the summary statistics $\boldsymbol{\Phi}_{(\mu,\sigma^2)}$ are here known:
\begin{equation}
\Phi^1_{(\mu,\sigma^2)} \thicksim \mathpzc{G}\left(\mu, \frac{\sigma^2}{n} \right) \enspace \mathrm{and} \enspace \Phi^2_{(\mu,\sigma^2)} \thicksim \varGamma\left( \frac{n-1}{2} , \frac{2\sigma^2}{n-1} \right)
\label{eq:Gaussian_proba_Phi1_Phi2}
\end{equation}
where $\varGamma$ is the Gamma distribution parametrised by its shape and scale.

The hierarchical graphical representation of the simulator is shown in figure \ref{fig:BHM_Gaussian_model}. 

\begin{figure*}
\includegraphics[width=\textwidth]{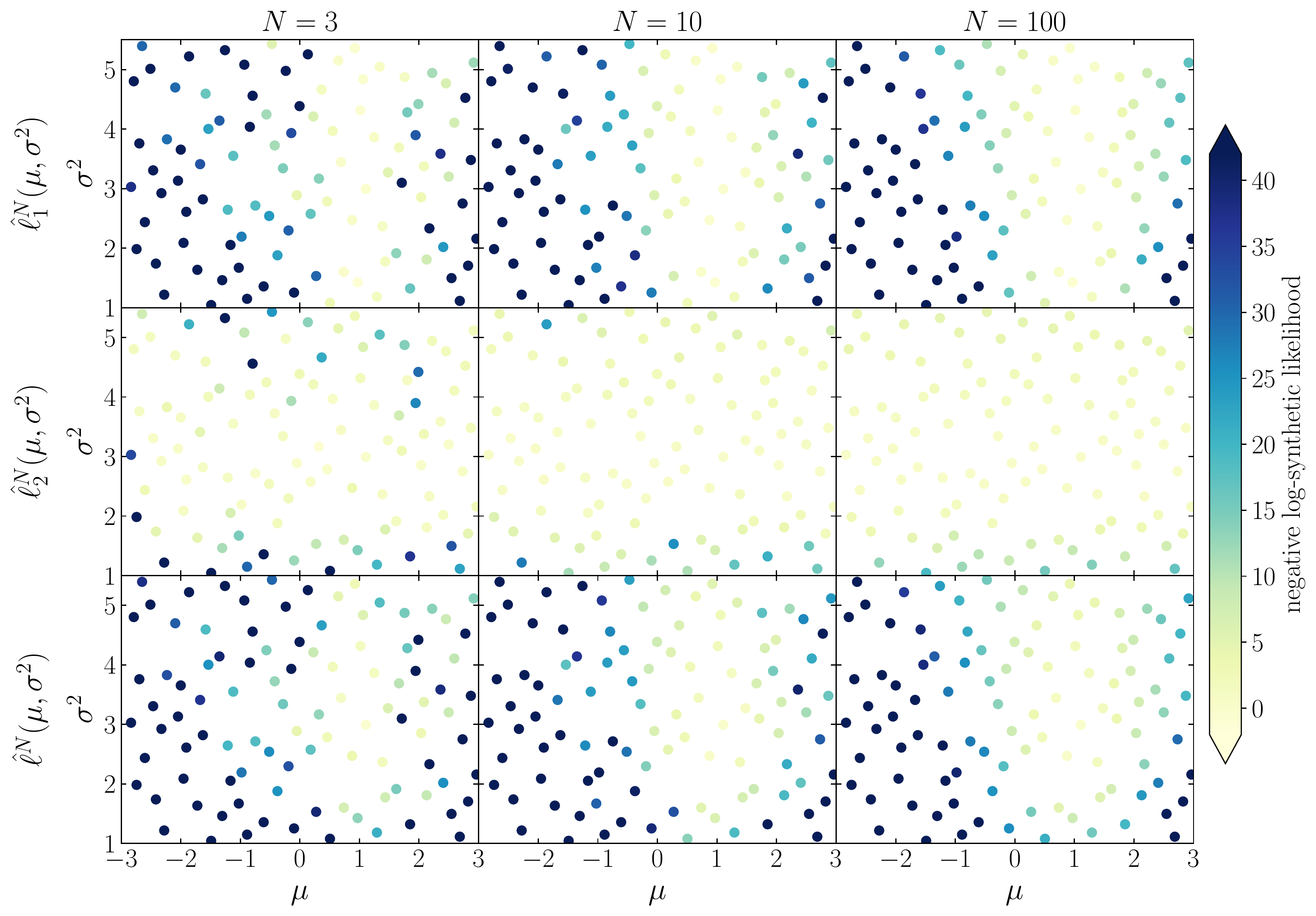}
\caption{Illustration of the Gaussian-Gamma synthetic likelihood as a stochastic process. The observed data have been generated using $\mu_\mathrm{true} = 0.8$ and $\sigma^2_\mathrm{true}=2.9$. The $100$ sampling points form a low-discrepancy quasi-random Sobol sequence in parameter space. The three rows show respectively the first term $\hat{\ell}_1^N(\mu,\sigma^2)$ (a Gaussian synthetic likelihood for $\Phi^1$), the second term $\hat{\ell}_2^N(\mu,\sigma^2)$ (a Gamma synthetic likelihood for $\Phi^2$), and their sum $\hat{\ell}^N(\mu,\sigma^2)$. The three columns show a varying number of simulations per value of $(\mu,\sigma^2)$: $N=3$, $N=10$, $N=100$. The use of simulations makes the synthetic likelihood a stochastic process. Its noisiness decreases as $N$ increases, i.e. as more computational resources are invested.\label{fig:Gaussian-Gamma_synthetic_likelihood}}
\end{figure*}

\subsection{Analytic solution}

The exact solution of the problem described in the previous section is known analytically: the posterior is Gaussian-inverse-Gamma distributed, with parameters $(\alpha', \beta', \eta', \lambda')$ given by
\begin{eqnarray}
\alpha' & = & \alpha + \frac{n}{2}, \label{eq:Gaussian_analytic_solution_alpha}\\
\beta' & = & \beta + \frac{n \lambda}{\lambda + n} \frac{( \Phi^1_\mathrm{O} - \eta)^2}{2} + \frac{n-1}{2} \Phi^2_\mathrm{O} , \label{eq:Gaussian_analytic_solution_beta} \\
\eta' & = & \frac{\lambda \eta + n \Phi^1_\mathrm{O}}{\lambda + n}, \label{eq:Gaussian_analytic_solution_eta} \\
\lambda' & = & \lambda + n, \label{eq:Gaussian_analytic_solution_lambda}
\end{eqnarray}
where $\Phi^1_\mathrm{O}$ and $\Phi^2_\mathrm{O}$ are the summary statistics of the observed data, defined by applying equations \eqref{eq:Gaussian_def_Phi1} and \eqref{eq:Gaussian_def_Phi2} to $\textbf{d}_\mathrm{O}$. 

For the experiment described in section \ref{sssec:Unknown mean, known variance}, we have used $n=10$ and $N=20$. The data have been generated from ground truth parameters $\mu_\mathrm{true} = 0.8$ and $\sigma^2_\mathrm{true} = 2.9$. We have measured $\Phi^1_\mathrm{O} = 1.3212$, and have chosen a Gaussian prior on $\mu$ with mean unity and variance unity. The exact posterior is therefore a Gaussian with mean $1.2490$ and variance $0.2248$.

For the experiment described in section \ref{sssec:Unknown mean and variance}, we have used $n=50$ and $N=10$. The data have been generated from ground truth parameters $\mu_\mathrm{true} = 0.8$ and $\sigma^2_\mathrm{true} = 2.9$ (shown as the plus in figure \ref{fig:Gaussian_mean_variance}). We have measured $\Phi^1_\mathrm{O} = 0.9925$ and  $\Phi^2_\mathrm{O} = 2.8499$. We have chosen a prior with parameters $(\alpha, \beta, \eta, \lambda) = (22,\, 54,\, 0,\, 6)$. The exact posterior has therefore parameters $(\alpha', \beta', \eta', \lambda') = (47,\, 127.8885,\, 0.8862,\, 56)$.

\subsection{Derivation of the Gaussian-Gamma synthetic likelihood for likelihood-free inference}
\label{sapx:Derivation of the Gaussian-Gamma synthetic likelihood for likelihood-free inference}

For likelihood-free inference, a computable approximation $\widehat{L}^N(\mu,\sigma^2)$ to the true likelihood given by equation \eqref{eq:Gaussian_true_likelihood} is required. In this section, we design a parametric form for $\widehat{L}^N(\mu,\sigma^2)$ which we call the Gaussian-Gamma synthetic likelihood.

As the approach is likelihood-free, $\widehat{L}^N(\mu,\sigma^2)$ should be based only on realisations of the summary statistics. Using the simulator described in section \ref{sapx:Forward modelling}, we can generate $N$ realisations of $\Phi^1$ and $\Phi^2$ for each pair of input parameters $(\mu,\sigma^2)$. Assuming exchangeability, we can use the Ansatz $L(\mu,\sigma^2) \equiv L_1(\mu,\sigma^2) \, L_2(\mu,\sigma^2)$ and $\widehat{L}^N(\mu,\sigma^2) \equiv \widehat{L}^N_1(\mu,\sigma^2) \, \widehat{L}^N_2(\mu,\sigma^2)$, or using the log-likelihood,
\begin{equation}
\hat{\ell}^N(\mu,\sigma^2) \equiv \hat{\ell}^N_1(\mu, \sigma^2) + \hat{\ell}^N_2(\mu, \sigma^2),
\end{equation}
where the first term depends only on $\Phi^1$ and the second on $\Phi^2$. They are discussed successively in the following.

$\Phi^1$ is the empirical mean of the independent and identically distributed components of $\textbf{d}$, obtained through averaging. As discussed in section \ref{sssec:Parametric approximations and the synthetic likelihood}, the Gaussian parametric approximation also known as the synthetic likelihood is appropriate in this case. We therefore define 
\begin{equation}
-2\hat{\ell}^N_1(\mu, \sigma^2) \equiv \log |2\pi \hat{v}^1_{(\mu,\sigma^2)}| + \frac{\left(\Phi^1_\mathrm{O} - \hat{\mu}^1_{(\mu,\sigma^2)} \right)^2}{\hat{v}^1_{(\mu,\sigma^2)}},
\end{equation}
where $\hat{\mu}^1_{(\mu,\sigma^2)}$ and $\hat{v}^1_{(\mu,\sigma^2)}$ are respectively the empirical mean and variance of the simulated $\Phi^1$, i.e.
\begin{eqnarray}
\hat{\mu}^1_{(\mu,\sigma^2)} & \equiv & \mathrm{E}^N\left[ \Phi^1_{(\mu,\sigma^2)} \right], \label{eq:Gaussian_def_mu_1} \\
\hat{v}^1_{(\mu,\sigma^2)} & \equiv & \mathrm{E}^N\left[ \left( \Phi^1_{(\mu, \sigma^2)} - \hat{\mu}^1_{(\mu,\sigma^2)} \right)^2 \right] . \label{eq:Gaussian_def_v_1}
\end{eqnarray}
As $\p(\Phi^1|\mu,\sigma^2)$ is actually a Gaussian distribution, the equality $\widetilde{L}_1(\mu,\sigma^2) = L_1(\mu,\sigma^2)$ holds without approximation, in the limit of infinite computer resources. From equation \eqref{eq:Gaussian_proba_Phi1_Phi2}, we also have
\begin{eqnarray}
\hat{\mu}^1_{(\mu,\sigma^2)} & \thicksim & \mathpzc{G}\left(\mu, \frac{\sigma^2}{Nn} \right) \quad \mathrm{and} \nonumber\\
\hat{v}^1_{(\mu,\sigma^2)} & \thicksim & \varGamma\left( \frac{N-1}{2} , \frac{2\sigma^2}{n(N-1)} \right),
\end{eqnarray}
which allows a closed-form definition of the stochastic process defining $\widehat{L}^N_1(\mu, \sigma^2)$.

$\Phi^2$ is the empirical variance of the components of $\textbf{d}$. As noted in equation \eqref{eq:Gaussian_proba_Phi1_Phi2}, $\p(\Phi^2|\mu,\sigma^2)$ is a Gamma distribution. Consequently, we introduce for $\Phi^2_\mathrm{O}$ a Gamma synthetic likelihood, namely
\begin{equation}
\begin{split}
-2\hat{\ell}^N_2(\mu,\sigma^2) \equiv &~-2(\hat{k}_{(\mu, \sigma^2)}-1)\log \Phi^2_\mathrm{O} + \frac{2\Phi^2_\mathrm{O}}{\hat{\theta}_{(\mu,\sigma^2)}} \\
& + 2\hat{k}_{(\mu, \sigma^2)} \log \hat{\theta}_{(\mu, \sigma^2)} + 2 \log \Gamma(\hat{k}_{(\mu, \sigma^2)}) .
\end{split}
\end{equation}
The question is now to use the simulator in order to learn the shape and scale parameters $\hat{k}_{(\mu, \sigma^2)}$ and $\hat{\theta}_{(\mu, \sigma^2)}$. To do so, the simplest possibility is the methods of moments: using a Gaussian approximation to the first two moments of the Gamma distribution, we have
\begin{eqnarray}
\hat{\mu}^2_{(\mu, \sigma^2)} & \approx & \hat{k}_{(\mu, \sigma^2)}\hat{\theta}_{(\mu, \sigma^2)}, \quad \mathrm{and} \\
\hat{v}^2_{(\mu, \sigma^2)} & \approx &  \hat{k}_{(\mu, \sigma^2)}\left( \hat{\theta}_{(\mu, \sigma^2)} \right)^2,
\end{eqnarray}
where $\hat{\mu}^2_{(\mu, \sigma^2)}$ and $\hat{v}^2_{(\mu, \sigma^2)}$ are the empirical mean and variance of $\Phi^2$, respectively, defined as in equations \eqref{eq:Gaussian_def_mu_1} and \eqref{eq:Gaussian_def_v_1}. Solving this system for $\hat{k}_{(\mu, \sigma^2)}$ and $\hat{\theta}_{(\mu, \sigma^2)}$, we obtain the parameters of $\hat{\ell}^N_2$,
\begin{eqnarray}
\hat{k}_{(\mu, \sigma^2)} & \approx & \frac{\left( \hat{\mu}^2_{(\mu, \sigma^2)} \right)^2}{\hat{v}^2_{(\mu, \sigma^2)}}, \quad \mathrm{and} \\
\hat{\theta}_{(\mu, \sigma^2)} & \approx & \frac{\hat{v}^2_{(\mu, \sigma^2)}}{\hat{\mu}^2_{(\mu, \sigma^2)}}.
\end{eqnarray}
As $\p(\Phi^2|\mu,\sigma^2)$ is known to be a Gamma distribution, we have, as for the first term, $\widetilde{L}^2(\mu,\sigma^2) = L^2(\mu,\sigma^2)$ in the limit of infinite computer resources. $\hat{\mu}^2_{(\mu,\sigma^2)}$ is the sum of $N$ independent random variables, identically distributed according to a Gamma distribution with the same scale parameter. Therefore, it obeys
\begin{equation}
\hat{\mu}^2_{(\mu,\sigma^2)} \thicksim \varGamma \left( \frac{N(n-1)}{2}, \frac{2\sigma^2}{N(n-1)} \right).
\end{equation}
Unlike $\hat{\mu}^2_{(\mu,\sigma^2)}$, there is no closed-form expression for $\hat{v}^2_{(\mu,\sigma^2)}$, $\hat{k}_{(\mu, \sigma^2)}$ and $\hat{\theta}_{(\mu, \sigma^2)}$ with standard probability distributions. However, these quantities, as well as $\widehat{L}_2^N(\mu,\sigma^2)$, can be easily simulated using their defining equations.

The resulting approximate likelihood $\widehat{L}^N(\mu,\sigma^2)$ is the product of a Gaussian synthetic likelihood for $\Phi^1$ and a Gamma synthetic likelihood for $\Phi^2$. It is shown in figure \ref{fig:Gaussian-Gamma_synthetic_likelihood}. There, the different panels show that realisations become smoother as $N$ increases, i.e. with more computational resources.

\section{Supernova cosmology}
\label{apx:Supernova cosmology}

This appendix gives the details of the data model and the modelling assumptions for the problem of inferring cosmological parameters from the JLA catalogue, presented in section \ref{ssec:Supernova cosmology}.

\subsection{Data samples}
\label{sapx:Data samples}

\begin{figure*}
\begin{center}
\includegraphics[width=\textwidth]{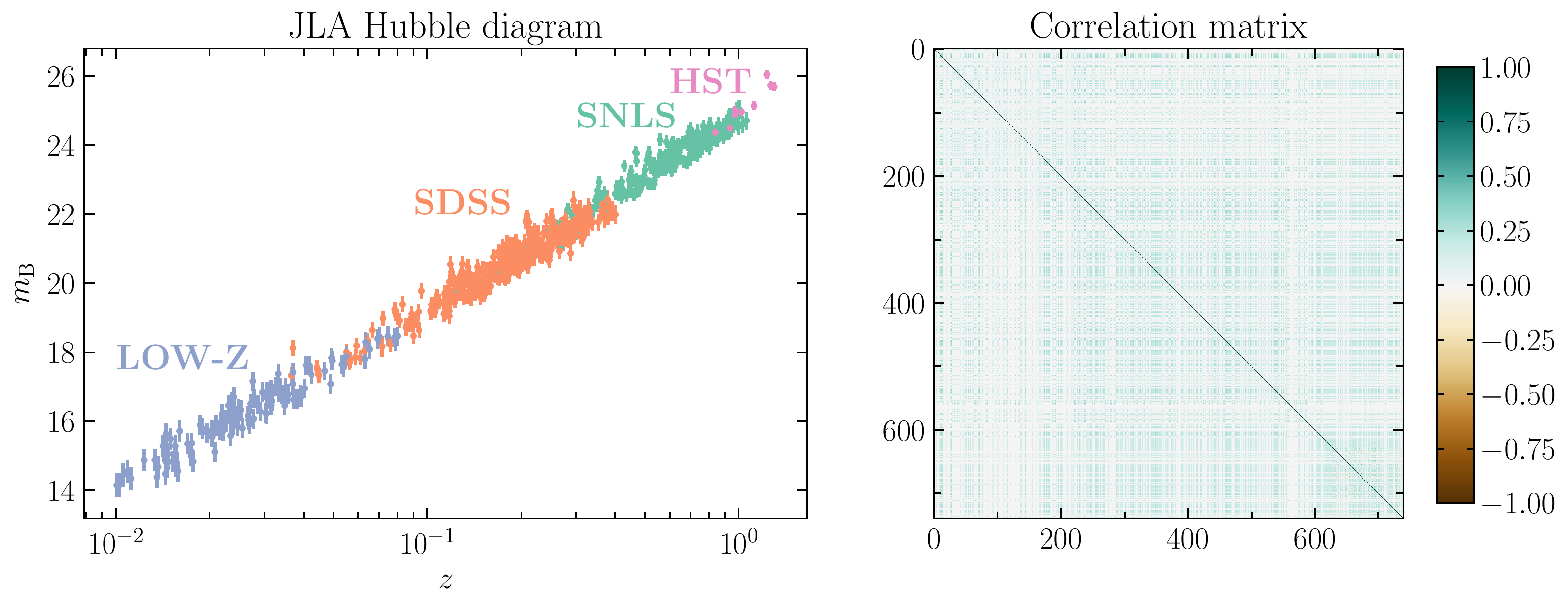} 
\caption{\textit{Left panel}. JLA Hubble diagram, representing the observed apparent magnitudes $m_\mathrm{B}$ of $740$ type Ia supernovae as a function of their redshift. The error bars represented correspond to $2 \Delta m_\mathrm{B}$, where $\Delta m_\mathrm{B}$ is included in the JLA catalogue but not used in this work. The different colours correspond to the different observational programmes used in the compilation. \textit{Right panel}. Correlation matrix of the observed apparent magnitudes, taking into account statistical and various systematic uncertainties \citep[see][section 5.5 for details on the construction of the covariance matrix]{Betoule2014}. \label{fig:JLA_Hubble_correlation}}
\end{center}
\end{figure*}

\begin{figure}
\begin{center}
\begin{tikzpicture}
	\pgfdeclarelayer{background}
	\pgfdeclarelayer{foreground}
	\pgfsetlayers{background,main,foreground}

	\tikzstyle{probability}=[draw, thick, text centered, rounded corners, minimum height=1em, minimum width=1em, fill=green!20]
	\tikzstyle{variabl}=[draw, thick, text centered, circle, minimum height=1em, minimum width=1em]
	\tikzstyle{hyperparam}=[draw, thick, text centered, circle, minimum height=1em, minimum width=1em, fill=yellow!20]

	\def\blockdist{0.7}
	\def\shiftdist{0.7}

    \node (omega) [hyperparam]
    {$\boldsymbol{\upomega}$};
    \path (omega.west)+(2*\blockdist,0) node (S) [hyperparam]
    {$\textbf{S}$};
    \path (omega.south)+(0.5,-\blockdist) node (Omegamwproba) [probability]
    {$\p(\Omega_\mathrm{m},w|\boldsymbol{\upomega}, \textbf{S})$};
    \path (Omegamwproba.south)+(-\blockdist,-\blockdist) node (Omegam) [variabl]
    {$\Omega_\mathrm{m}$};
    \path (Omegam.east)+(\blockdist,0) node (w) [variabl]
    {$w$};
    \path (w.east)+(2*\blockdist,0) node (M) [hyperparam]
    {$\textbf{M}$};
    \path (M.east)+(\blockdist,0) node (m_O) [hyperparam]
    {$\textbf{m}_\mathrm{O}$};
   	\path (Omegam.south)+(3*\blockdist,-\blockdist) node (dproba) [probability]
	{$\p(\textbf{d}|\Omega_\mathrm{m},w,\textbf{M},\textbf{m}_\mathrm{O})$};
   	\path (dproba.south)+(0,-\blockdist) node (d) [variabl]
    {$\textbf{d}$};

	\draw[line width=0.7pt, arrows={-latex}] (omega) -- (Omegamwproba);
	\draw[line width=0.7pt, arrows={-latex}] (S) -- (Omegamwproba);
	\draw[line width=0.7pt, arrows={-latex}] (Omegamwproba) -- (Omegam);
	\draw[line width=0.7pt, arrows={-latex}] (Omegamwproba) -- (w);
	\draw[line width=0.7pt, arrows={-latex}] (Omegam) -- (dproba);
	\draw[line width=0.7pt, arrows={-latex}] (w) -- (dproba);
	\draw[line width=0.7pt, arrows={-latex}] (M) -- (dproba);
	\draw[line width=0.7pt, arrows={-latex}] (m_O) -- (dproba);
	\draw[line width=0.7pt, arrows={-latex}] (dproba) -- (d);

\end{tikzpicture}
\end{center}
\caption{Hierarchical forward model for the analysis of the JLA type Ia supernovae catalogue. The prior on the physical parameters $\Omega_\mathrm{m}$ and $w$ is a Gaussian with mean $\boldsymbol{\upomega}$ and covariance matrix $\textbf{S}$. The data generating process uses four nuisance parameters, the distribution of which is characterised by the hyperparameters $\textbf{M}$ and the supernovae metadata $\textbf{m}_\mathrm{O}$.\label{fig:BHM_supernovae}}
\end{figure}
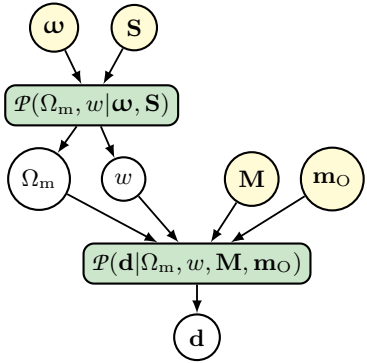

Type Ia supernovae (SNe Ia) are ``standard candles'', i.e. astrophysical objects that precisely map the distance-redshift relation in the nearby Universe. As such, they are one of the most sensitive probes of the late-time expansion history of the Universe. The Joint Lightcurve Analysis \citep[JLA,][]{Betoule2014} is a compiled catalogue of $740$ SNe Ia. $374$ objects in the redshift range $0.03 \leq z \leq 0.41$ have been identified by the Sloan Digital Sky Survey phase II (SDSS-II) supernova survey \citep{Sako2018} and confirmed as SNe Ia by spectroscopic follow-up observations. The remaining objects come from the earlier C11 compilation \citep{Conley2011}: $118$ are low-$z$ ($z \leq 0.08$) SNe Ia from the third release \citep{Hicken2009} of photometric data acquired at the Whipple Observatory of the Harvard-Smithsonian Center for Astrophysics (CfA3). $239$ SNe Ia in the redshift range $0.12 \leq z \leq 1.07$ have been observed by the Supernova Legacy Survey \citep[SNLS,][]{Astier2006,Sullivan2011}. Finally, $9$ objects are high-redshift SNe Ia ($0.8 \leq z \leq 1.4$) observed by the Hubble Space Telescope \citep[HST,][]{Riess2007}.

For each supernova, the JLA catalogue provides a rich variety of information. The full data set comprises lightcurves in different bands and spectroscopic or photometric observations of each SN Ia. These products are then used to estimate the redshift $z$, the apparent magnitude $m$, the colour at maximum brightness $C$ and a time-stretching parameter for the lightcurve, $X_1$. In particular, the catalogue includes several estimations of the redshift $z$. In this work, we use $z = z_\mathrm{CMB}$, the cosmological redshift of the object in the frame of the cosmic microwave background (CMB), including peculiar velocity corrections. For our data vector $\textbf{d}_\mathrm{O}$, we use the estimated B-band peak magnitudes in the rest frame, denoted $\left( m_\mathrm{B,O}^k \right)$ for $k \in \llbracket 1, 740 \rrbracket$ (as in the body of the paper, the subscript $\mathrm{O}$ stands for ``observed''). The magnitudes are plotted as a function of redshift in the Hubble diagram shown in figure \ref{fig:JLA_Hubble_correlation} (left). The JLA catalogue also provides some properties of the SNe host galaxies, in particular the stellar mass $M_\mathrm{stellar}$. We denote by $\textbf{z}_\mathrm{O} \equiv \left( z^k_\mathrm{O} \right)$, $\textbf{X}_{1,\mathrm{O}} \equiv \left( X_{1,\mathrm{O}}^k \right)$, $\textbf{C}_\mathrm{O} \equiv \left( C^k_\mathrm{O} \right)$, $\textbf{M}_\mathrm{stellar, O} \equiv \left( M_\mathrm{stellar,O}^k \right)$ for $k \in \llbracket 1, 740 \rrbracket$, and $\textbf{m}_\mathrm{O} \equiv \left( \textbf{z}_\mathrm{O}, \textbf{X}_{1,\mathrm{O}}, \textbf{C}_\mathrm{O}, \textbf{M}_\mathrm{stellar, O} \right)$ the metadata used in the analysis.

\subsection{Supernova data model and distance estimates}
\label{sapx:Supernova data model and distance estimates}

\begin{figure*}
\begin{center}
\includegraphics[width=\textwidth]{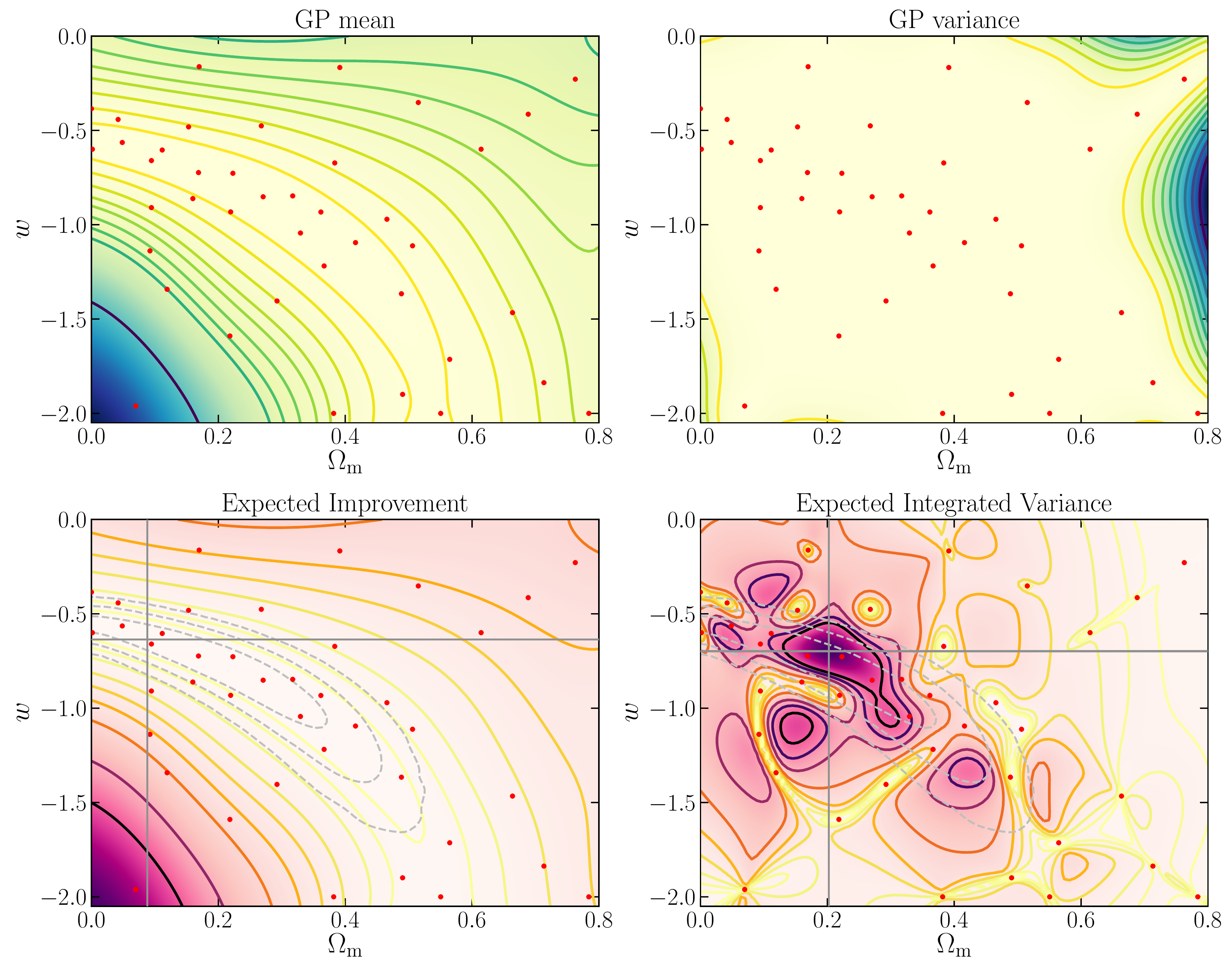} 
\caption{{\bolfi} at work after $20$ acquisitions for the supernovae cosmology problem. \textit{Top panels}. Isocontours of the Gaussian process model for the discrepancy $\Delta_{(\Omega_\mathrm{m},w)}$. The mean (left) and variance (right) are shown in arbitrary units. The red dots mark the location of the training parameters $(\Omega_\mathrm{m},w)$. \textit{Bottom panels}. Isocontours of the acquisition surfaces built from the Gaussian process, using two different acquisition rules: the expected improvement (which is maximised, left), and the expected integrated variance (which is minimised, right). Units are arbitrary. The location of the next acquisition (i.e. the optimiser) is marked by the cross, and the contours of the exact posterior are plotted as dashed gray lines for reference. The initial training set is composed of $20$ samples, and the expected integrated variance has been used for the $20$ acquisitions shown.\label{fig:Supernovae_acquisition}}
\end{center}
\end{figure*}

Distance estimation with SNe Ia is based on the assumption that they are standardizable objects, which is quantified by a linear model for the apparent magnitude:
\begin{equation}
m_\mathrm{B} = 5 \log_{10} \left[ \frac{D_\mathrm{L}(z)}{10~\mathrm{pc}} \right] + \widetilde{M}_\mathrm{B}(M_\mathrm{stellar}, M_\mathrm{B}, \deltaM) - \alpha X_1 + \beta C.
\label{eq:SNe_model_distance}
\end{equation}
The absolute magnitude $\widetilde{M}_\mathrm{B}$ depends on the stellar mass of the host galaxy, $M_\mathrm{stellar}$. This dependence is assumed to be captured by the relation \citep{Conley2011}
\begin{equation}
\widetilde{M}_\mathrm{B}(M_\mathrm{stellar}, M_\mathrm{B}, \deltaM) = M_\mathrm{B} + \deltaM \, \Theta\left( M_\mathrm{stellar} - 10^{10} \mathrm{M}_\odot \right),
\label{eq:SNe_model_magnitude}
\end{equation}
where $\Theta$ is the Heaviside function and $\mathrm{M}_\odot$ the mass of the Sun. The lightcurve calibration model therefore comprises four nuisance parameters ($\alpha$, $\beta$, $M_\mathrm{B}$, $\deltaM$). They are assumed to be independent of host galaxy properties. 

The cosmological model enters in the analysis through the distance-redshift relation. We assume a flat Universe containing cold dark matter and a dark energy component ($w$CDM hereafter). A $w$CDM Universe is characterised by two physical parameters $\Omega_\mathrm{m}$ (the matter density) and $w$ (the equation of state of dark energy, assumed constant in time). The luminosity distance appearing in equation \eqref{eq:SNe_model_distance} is given by \citep[e.g.][section 7]{Hogg1999}
\begin{eqnarray}
D_\mathrm{L}(z) & = & \frac{(1+z) \, \mathrm{c}}{H_0} \int_0^z \frac{\mathrm{d}z'}{E(z')}, \nonumber\\
E(z) & \equiv & \sqrt{\Omega_\mathrm{m}(1+z)^3 + (1-\Omega_\mathrm{m})(1+z)^{3(w+1)}}, \quad \label{eq:luminosity_distance_def}
\end{eqnarray}
where $\mathrm{c}$ is the speed of light in vacuum and $H_0 \equiv 100 \, h~\mathrm{km}\,\mathrm{s}^{-1}\,\mathrm{Mpc}^{-1}$.

\subsection{Forward modelling}

The data model described in the previous section can be simulated forward by taking the following operations successively:
\begin{eqnarray}
(\Omega_\mathrm{m},w) & \curvearrowleft & \p(\Omega_\mathrm{m},w | \boldsymbol{\upomega}, \textbf{S}), \label{eq:SNe_forward_1}\\
(\alpha, \beta, M_\mathrm{B}, \deltaM) & \curvearrowleft & \p(\alpha, \beta, M_\mathrm{B}, \deltaM | \textbf{M}), \label{eq:SNe_forward_2}\\
D_\mathrm{L}(\textbf{z}_\mathrm{O}) & \curvearrowleft & \p(D_\mathrm{L}(\textbf{z}_\mathrm{O}) | \Omega_\mathrm{m}, w), \label{eq:SNe_forward_3}\\
\textbf{d} & \curvearrowleft & \p(\textbf{d} | D_\mathrm{L}(\textbf{z}_\mathrm{O}), \alpha, \beta, M_\mathrm{B}, \deltaM, \textbf{m}_\mathrm{O}). \quad\quad\label{eq:SNe_forward_4}
\end{eqnarray}
The last two steps are deterministic: in equation \eqref{eq:SNe_forward_3}, the luminosity distance at the observed redshifts is computed via equation \eqref{eq:luminosity_distance_def}, and in equation \eqref{eq:SNe_forward_4}, the predicted data $\textbf{d}_{(\Omega_\mathrm{m},w)} \equiv \left( m_{\mathrm{B},(\Omega_\mathrm{m},w)}^k \right)$ come from equations \eqref{eq:SNe_model_distance} and \eqref{eq:SNe_model_magnitude}. We can therefore write
\begin{eqnarray}
\begin{aligned}
& \begin{split}
& \p(D_\mathrm{L}(\textbf{z}_\mathrm{O})| \Omega_\mathrm{m}, w) \\
& \quad = \updelta_\mathrm{D}\left( D_\mathrm{L}(\textbf{z}_\mathrm{O}) - \widehat{D}_\mathrm{L}(\textbf{z}_\mathrm{O}, \Omega_\mathrm{m}, w)\right) ,\end{split}\\
& \begin{split}
& \p(\textbf{d} | D_\mathrm{L}(\textbf{z}_\mathrm{O}), \alpha, \beta, M_\mathrm{B}, \deltaM, \textbf{m}_\mathrm{O}) \\
& \quad = \updelta_\mathrm{D}\left( \textbf{d} -\boldsymbol{\hat{\mathrm{d}}}(D_\mathrm{L}(\textbf{z}_\mathrm{O}), \alpha, \beta, M_\mathrm{B}, \deltaM, \textbf{m}_\mathrm{O})\right), \end{split}\\
& \begin{split}
& \p(\textbf{d} | \Omega_\mathrm{m},w, \textbf{M}, \textbf{m}_\mathrm{O}) \\
& \quad = \updelta_\mathrm{D}\left( \textbf{d} -\boldsymbol{\hat{\mathrm{d}}}(D_\mathrm{L}(\textbf{z}_\mathrm{O}), \alpha, \beta, M_\mathrm{B}, \deltaM, \textbf{m}_\mathrm{O})\right) \\
& \quad\quad \times \updelta_\mathrm{D}\left( D_\mathrm{L}(\textbf{z}_\mathrm{O}) - \widehat{D}_\mathrm{L}(\textbf{z}_\mathrm{O}, \Omega_\mathrm{m}, w)\right) \\
& \quad\quad \times \p(\alpha, \beta, M_\mathrm{B}, \deltaM | \textbf{M}). \end{split}
\end{aligned}
\end{eqnarray}

The probability $\p(\Omega_\mathrm{m},w | \boldsymbol{\upomega}, \textbf{S})$ appearing in equation \eqref{eq:SNe_forward_1} is the Gaussian prior given by equation \eqref{eq:SNe_prior_Omegam_w}, i.e. $\p(\Omega_\mathrm{m},w | \boldsymbol{\upomega}, \textbf{S}) \equiv \mathpzc{G}(\boldsymbol{\upomega}, \textbf{S})$ with
\begin{equation}
\boldsymbol{\upomega} \equiv \begin{pmatrix}
0.3 \\
0.75
\end{pmatrix} \quad \mathrm{and} \quad
\textbf{S} \equiv \begin{pmatrix}
0.4^2 & -0.24 \\
-0.24 & 0.75^2
\end{pmatrix}.
\end{equation}

Finally, $\p(\alpha, \beta, M_\mathrm{B}, \deltaM | \textbf{M})$ is the sampling distribution of nuisance parameters, characterised by hyperparameters $\textbf{M}$. Following previous studies, we choose broad, independent Gaussian priors on each of the four parameters. Specifically, we assume
\begin{equation}
\begin{pmatrix}
\alpha \\
\beta \\
M_\mathrm{B} \\
\deltaM
\end{pmatrix} \sim \mathpzc{G} \left[
\begin{pmatrix}
0.125 \\
2.6 \\
-19.05 \\
-0.05
\end{pmatrix},
\begin{pmatrix}
0.025^2 & 0 & 0 & 0\\
0 & 0.25^2 & 0 & 0\\
0 & 0 & 0.1^2 & 0\\
0 & 0 & 0 & 0.03^2
\end{pmatrix} \right].
\end{equation}

The hierarchical graphical representation of the simulator is shown in figure \ref{fig:BHM_supernovae}.

\subsection{Discrepancy}
\label{sapx:Discrepancy}

Following \citet[][formula 15]{Betoule2014}, we define the discrepancy between observed and simulated data as
\begin{equation}
\Delta_{(\Omega_\mathrm{m},w)} \equiv (\textbf{d}_\mathrm{O} - \boldsymbol{\hat{\upmu}}_{(\Omega_\mathrm{m},w)})^\intercal \textbf{C}^{-1} (\textbf{d}_\mathrm{O} - \boldsymbol{\hat{\upmu}}_{(\Omega_\mathrm{m},w)}),
\label{eq:Supernovae_discrepancy}
\end{equation}
where $\boldsymbol{\hat{\upmu}}_{(\Omega_\mathrm{m},w)}$ is the average of $N$ simulated realisations of $\textbf{d}_{(\Omega_\mathrm{m},w)} \equiv \left( m_{\mathrm{B},(\Omega_\mathrm{m},w)}^k \right)$ for $k \in \llbracket 1,740 \rrbracket$. This is equivalent to assuming a Gaussian synthetic likelihood (see section \ref{sssec:Parametric approximations and the synthetic likelihood}) in approximate Bayesian computation, and to using a Gaussian likelihood for the exact Bayesian problem, solved by MCMC sampling for reference. \citet[][section 5.5]{Betoule2014} constructed a covariance matrix $\textbf{C}_{(\alpha,\beta)}$ which accounts for the uncertainty in the colour, stretch and redshift of each supernova, depending on the nuisance parameters $\alpha$ and $\beta$, but dropped the term $\log |2\pi\textbf{C}_{(\alpha,\beta)}|$ from the definition of the discrepancy. Since $\alpha$ and $\beta$ are very well constrained by the data, the dependence of $\textbf{C}_{(\alpha,\beta)}$ has a weak effect on the final inference results. Therefore, in this work \citep[and as in][]{Alsing2018}, we assume a fixed covariance matrix $\textbf{C}$ where the parameters $\alpha$ and $\beta$ are taken at their maximum \textit{a posteriori} value ($\alpha = 0.1256$, $\beta=2.6342$). This also justifies dropping the constant term $\log |2\pi \textbf{C}|$ from the definition of the discrepancy.

We used the data (version 6) and the python script provided along with the JLA\footnote{These products are available at \href{http://supernovae.in2p3.fr/sdss_snls_jla/ReadMe.html}{http://supernovae.in2p3.fr/sdss\_snls\_jla/ReadMe.html}.} to generate the $740 \times 740$ covariance matrix $\textbf{C}$. The associated correlation matrix is shown in figure \ref{fig:JLA_Hubble_correlation} (right).

\subsection{Acquisition}

For the analysis described in section \ref{ssec:Supernova cosmology}, we used $N=50$ simulations per point $(\Omega_\mathrm{m},w)$, and the ExpIntVar rule without acquisition noise. Figure \ref{fig:Supernovae_acquisition} shows the state of {\bolfi} after $20$ acquisitions, for a training set of $40$ samples. As can be observed in the lower panels, the different acquisition functions implement a different trade-off between exploration and exploitation. In particular, the ExpIntVar surface has a much more complex structure. Simulations surrounding the $3\sigma$ contour of the posterior have already been run (exploration). The proposed acquisition is in a region of high estimated density (exploitation), but not yet fully sampled. On the contrary, the next acquisition suggested by the EI criterion stays in the ``valley'' (the innermost contour line) where lies the estimated optimum, meaning that the tails of the posterior will hardly be sufficiently sampled.

\acknowledgments

\small{The author thanks Jens Jasche and Wolfgang Enzi for the collaboration that triggered this project, and Alan Heavens for useful discussions and a careful reading of the manuscript. This work has made use of a modified version of the \textsc{elfi} \citep[Engine for Likelihood-Free Inference,][]{Lintusaari2017b} code. The author acknowledges funding from the Imperial College London Research Fellowship Scheme.}

\renewcommand{\emph}[1]{\textit{#1}}

\section*{References}
\bibliography{/home/leclercq/workspace/biblio/biblio}

\end{document}